\documentclass[aps,prl,reprint,showpacs,showkeys,superscriptaddress,preprintnumbers, secnumarabic,nofootinbib,tightenlines,nobibnotes]{revtex4-1}
\usepackage{amsmath,amssymb,bm,mathrsfs}
\usepackage{srcltx,hyperref}
\usepackage{epsfig}
\usepackage{slashed}
\usepackage{bbold}
\usepackage{psfrag}
\usepackage{color}
\PassOptionsToPackage{caption=false}{subfig}
\usepackage{subfig}
\usepackage{multirow}
\usepackage{booktabs}
\usepackage{hyperref}

\newcommand{\ccbar}{{c\bar{c}}}
\newcommand{\bbbar}{{b\bar{b}}}

\newcommand{\eg}{{\it e.g.}}
\newcommand{\beq}{\begin{equation}}
\newcommand{\eeq}{\end{equation}}

\newcommand{\cR}{\mathcal{R}}
\newcommand{\BR}{{\rm BR}}

\begin{document}

\preprint{KEK-TH-1813}

\title{Constraining the Charm Yukawa and Higgs--quark Coupling Universality}

\author{Gilad Perez}
\email{gilad.perez@weizmann.ac.il}
\affiliation{Department of Particle Physics and Astrophysics,
Weizmann Institute of Science, Rehovot 7610001, Israel} 
\author{Yotam Soreq}
\email{yotam.soreq@weizmann.ac.il}
\affiliation{Department of Particle Physics and Astrophysics,
Weizmann Institute of Science, Rehovot 7610001, Israel} 

\author{Emmanuel Stamou}
\email{emmanuel.stamou@weizmann.ac.il}
\affiliation{Department of Particle Physics and Astrophysics,
Weizmann Institute of Science, Rehovot 7610001, Israel} 

\author{Kohsaku Tobioka}
\email{kohsakut@post.tau.ac.il}
\affiliation{Department of Particle Physics and Astrophysics,
Weizmann Institute of Science, Rehovot 7610001, Israel} 
\affiliation{Theory Center, High Energy Accelerator Research Organization (KEK), Tsukuba 305-0801, Japan}
\affiliation{School of Physics and Astronomy, Raymond and Beverley Sackler Faculty of Exact Sciences, Tel Aviv University, Tel Aviv 6997801, Israel}

\begin{abstract}
We introduce four different types of data-driven analyses with different level of 
robustness that constrain the size of the Higgs-charm Yukawa coupling:
(i)~recasting the vector-boson associated, $Vh$, analyses that search for the 
bottom-pair final state. 
We use this mode to directly and model independently constrain the Higgs to charm 
coupling, $y_c/y_c^{\rm SM}\lesssim 234$;
(ii)~the direct measurement of the total width, $y_c/y_c^{\rm SM}\lesssim 120{\rm -}140$;
(iii)~the search for $h\to J/\psi\gamma$, $y_c/y_c^{\rm SM}\lesssim 220$; 
(iv)~a global fit to the Higgs signal strengths, 
$y_c/y_c^{\rm SM}\lesssim 6.2\,.$
A comparison with $t\bar{t}h$ data allows us to show that the Higgs does not couple to quarks in a universal way, 
as is expected in the Standard Model. 
Finally, we demonstrate how the experimental collaborations can further improve our 
direct bound by roughly an order of magnitude by charm-tagging as already used in 
new-physics searches.
\end{abstract}

\maketitle

{\bf Introduction: } 
The discovery of the Higgs boson is a triumph of the LHC~\cite{Aad:2012tfa,Chatrchyan:2012ufa} 
and yet another success for the Standard Model~(SM) with its minimal Higgs sector of 
electroweak~(EW) symmetry breaking~(EWSB). 
The first run of the LHC was very successful not only because of the Higgs discovery, 
but also because it provided us with a rather strong qualitative  test of several aspects of the Higgs mechanism: it established that the 
Higgs plays a dominant role in inducing the masses of the EW gauge bosons and that 
the Higgs coupling to the longitudinal states tames the $WW$ scattering rates 
up to high energies. 

However, in the minimalistic SM way of EWSB the Higgs plays another
crucial role. Namely, it induces the masses of all charged fermions. 
This results in a sharp prediction, free of additional input parameters, for 
the Higgs--fermion interaction strength
\begin{equation}
y_f \simeq \sqrt2 \,{m_f\over v}\,,
\end{equation}  
where $f=u,c,t,d,s,b,e,\mu,\tau$ and $v\simeq246\,$GeV is the Higgs vacuum expectation value. 
This prediction holds to a very good accuracy. 
So far, this additional role of the Higgs has not yet 
been tested directly in a strong way.
The best information currently available is on the Higgs couplings to the third-generation 
charged fermions
\begin{equation} \label{eq:3rdavg} 
	\hspace*{-.3cm}	
	\mu_{t\bar t h}= 2.4\pm0.8,  \,
	\mu_{b}=0.71\pm0.31,   \,
	\mu_{\tau}= 0.98\pm0.22\,. 		
\end{equation}
Here, we averaged the ATLAS~\cite{ATLAStth,Aad:2014xzb,Aad:2015vsa} and 
CMS~\cite{Khachatryan:2014qaa,Chatrchyan:2013zna,Chatrchyan:2014nva} results for the 
Higgs signal strength to fermions $\mu_{f}\equiv \frac{\sigma}{\sigma_{\rm SM}}\,\frac{\BR_{f\bar f}}{\BR_{ f\bar f}^{\rm SM}}$. $\sigma$ stands for the production cross section, 
$\BR_{X}=\BR(h\to X)$ and the SM script indicates the SM case. 
These results are consistent with the SM expectations, though the errors are still 
noticeably large.
In contrast, our current knowledge regarding the Higgs couplings to the first 
two light generation fermions, is significantly poorer. 
In fact, at this point we only have a rather weak upper bound on the corresponding 
signal strengths of muons and electrons~\cite{Aad:2014xva,Khachatryan:2014aep} 
\begin{equation} \label{eq:muemu}
	\mu_{\mu}\leq 7 \, ,  \qquad
	\mu_{e}\leq 4 \times 10^5 \,,
\end{equation}
at 95\%~Confidence Level (CL). Eqs.~\eqref{eq:3rdavg} and \eqref{eq:muemu} 
together exclude Higgs--lepton universality.
Direct information does not exist at present regarding the Higgs--light-quark 
couplings. 
Measuring these Higgs--light couplings is interesting for three reasons. 
The first, although somewhat mundane, is simply that the light-quark 
Yukawa couplings are parameters of the SM and as such merit a measurement.
The second is that given the success of both direct and indirect tests of the SM 
it is now expected that the EW gauge bosons and the top quark acquire their masses
dominantly via the Higgs mechanism; this is less obvious for the first two 
generation quarks. 
The light-quark masses could be induced by other subdominant sources of EWSB, for instance 
from a technicolor-like condensate. Hence, light quarks may have suppressed or even vanishing 
Yukawa couplings to the Higgs. 
In fact, based on current knowledge, we could just add bare mass terms to the first two 
generation fermions and treat the SM as an effective theory that is valid up to some 
fairly high scale, at which ``unitarity'' or the weakly-coupled description would breakdown. 
This is similar to the status of the EW gauge sector prior to the first run of the LHC. 
If we assume no coupling of light quarks to the Higgs, the unitarity bound from
the $q\bar{q}\to V_L V_L$ process (where $V_L$ is a longitudinal boson) is (see {\it e.g.} 
Refs.~\cite{Chivukula:2007mw, Marciano:1989ns, Appelquist:1987cf}) 
\begin{eqnarray}
	\sqrt{s} 
	&\lesssim& \frac{8\pi v^2}{\sqrt{6} m_{b,c,s,d,u}}  \nonumber \\
	&\approx& 200,\, 1\!\times\! 10^3,\, 1\!\times\!10^4,\, 2\!\times\!10^5,\, 5\!\times\! 10^5\,\,{\rm TeV}\, .
\label{eq:ouracresult}
\end{eqnarray}
Even stronger bounds are found when $q\bar{q}\to n V_L$ processes are 
considered \cite{Maltoni:2000iq}. The lead to the following 
corresponding unitarity constraints \cite{Dicus:2004rg}, 
%
\begin{eqnarray}
	\sqrt{s} 
	\lesssim 23,\, 31 ,\, 52 ,\, 77,\, 84\,\,{\rm TeV}\, .
\label{eq:ouracresult}
\end{eqnarray}
These bounds are weak enough as to make the question regarding the origin
of light-quark masses a fundamentally interesting question. 
The third argument, following a reverse reasoning, is that with new physics it 
is actually easy to obtain enhancements in Higgs--light-quark interaction 
strengths. 
As the Higgs is rather light it can only decay to particles that interact very weakly with it. 
Within the SM, its dominant decay mode is to bottom quark pair. 
Therefore, a deformation of the Higgs couplings to the lighter SM particles, 
say the charm quarks (for possibly relevant discussions see 
Ref.~\cite{Delaunay:2013iia,Delaunay:2013pwa,Blanke:2013uia,Mahbubani:2012qq,Kagan:2009bn,Dery:2013aba,Giudice:2008uua,DaRold:2012sz,Chen:2013qta,Dery:2014kxa}), 
could compete with the Higgs--bottom coupling and would lead to a dramatic change of the 
Higgs phenomenology at collider~\cite{Delaunay:2013pja}.

Recent theoretical and experimental progress opened a window towards studying the Higgs 
coupling to light quarks at future colliders. 
On the theoretical frontier, it was demonstrated in Ref.~\cite{Delaunay:2013pja} that 
using inclusive charm-tagging would enable the LHC experiments to search for the decay 
of the Higgs into a pair of charm jets ($c$-jets). 
Furthermore, it was shown that the Higgs--charm coupling may be probed by looking at 
exclusive decay modes involving a $c$-$\bar c$ vector meson and a photon~\cite{Bodwin:2013gca}. 
A similar mechanism, based on exclusive decays to light-quark states and gauge bosons $\gamma/W/Z$, 
was shown to yield a potential access to the Higgs--light-quark couplings~\cite{Kagan:2014ila}. 
(See also Refs.~\cite{Mangano:2014xta,Huang:2014cxa, Grossmann:2015lea} for studies 
of exclusive EW gauge boson decays.) 
On the experimental frontier, ATLAS has recently published two SUSY 
searches \cite{Aad:2014nra, Aad:2015gna} that make use of charm-tagging \cite{ATL-PHYS-PUB-2015-001}.  
On the exclusive frontier, ATLAS searched for Higgs decays
to quarkonia({\it e.g.} $J/\psi$, $\Upsilon$) and a photon final state~\cite{Aad:2015sda}. 
All these developments provide a proof of principle that in the future we may 
be able to test the Higgs mechanism of mass generation even for light quarks.  

In the following we introduce four different types of data-driven analyses 
with different level of robustness that constrain the size of the Higgs--charm 
Yukawa coupling. 
This should be considered as a first step towards improving our understanding 
regarding the origin of light-quark masses.
In the future, the methods described below are expected to yield significantly better sensitivities 
to the corresponding Yukawa couplings. 
One direct implication of our analyses is the establishment of the fact that the Higgs 
couples to the quarks in a non-universal manner.

{\bf Signal-strength constraint via $\boldsymbol{Vh(b\bar b)}$ recast:}
The ATLAS and CMS collaborations studied the Higgs decay into $b\bar{b}$ via 
$Vh$ production, in which the Higgs is produced in association with a $W/Z$ gauge boson, 
using $5\,$fb$^{-1}$\,at $7\,$TeV and $20\,$fb$^{-1}$\,at $8\,$TeV~\cite{Aad:2014xzb,Chatrchyan:2013zna}. 
Due to the rough similarities between charm and bottom jets, jets originating from 
charm quarks may be mistagged as $b$-jets. 
We thus recast the existing analyses of $h\to b\bar{b}$ to study and constrain 
the $h\to c\bar{c}$ rate. 
This will provide a direct and model-independent bound on the Higgs--charm coupling.
To allow the Higgs--charm coupling to float freely, the signal strength should be modified
according to
\begin{equation}
\begin{split}
	\mu_b&= \frac{\sigma\, \BR_{\bbbar} 
	}
	{\sigma_{\rm SM}  \BR^{\rm SM}_{\bbbar}
	}\\
	&\to 
	\frac{\sigma\, \BR_{\bbbar} \, \epsilon_{b_1}\epsilon_{b_2}
	+\sigma\, \BR_{\ccbar} \, \epsilon_{c_1}\epsilon_{c_2}}
	{\sigma_{\rm SM}  \BR^{\rm SM}_{\bbbar}\, \epsilon_{b_1}\epsilon_{b_2}
	+\sigma_{\rm SM}\, \BR^{\rm SM}_{\ccbar} \, \epsilon_{c_1}\epsilon_{c_2}}
	\\
	&=\left(\mu_b +\frac{\BR_{\ccbar}^{\rm SM}}{\BR_{\bbbar}^{\rm SM}}
	\frac{\epsilon_{c_1}\epsilon_{c_2}}{\epsilon_{b_1}\epsilon_{b_2}} \mu_c \right)\!\!\Bigg/\!\!
	\left(1 +\frac{\BR_{\ccbar}^{\rm SM}}{\BR_{\bbbar}^{\rm SM}}
	\frac{\epsilon_{c_1}\epsilon_{c_2}}{\epsilon_{b_1}\epsilon_{b_2}} \right)
	,
\end{split}
\label{mubc}
\end{equation}
where $\epsilon_{b_{1,2}}$ and $\epsilon_{c_{1,2}}$ are efficiencies to tag 
jets originating from bottom and charm quarks, respectively, and 
$\BR_{\ccbar}^{\rm SM}/\BR_{\bbbar}^{\rm SM}\simeq5\%$~\cite{Heinemeyer:2013tqa}.

One working point for $b$-tagging and $c$-jet contamination, 
defined via $\epsilon_{b_{1,2}},\epsilon_{c_{1,2}}$, constrains only one linear 
combination of $\mu_b$ and $\mu_c$; it corresponds to a flat direction in the  $\mu_c$--$\mu_b$ plane. 
To disentangle the flat direction, at least two tagging points with different 
ratios, $\epsilon^2_{c/b}\equiv  (\epsilon_{c_{1}}\epsilon_{c_{2}})/(\epsilon_{b_{1}}\epsilon_{b_{2}})$, 
should be adopted. 
Both ATLAS and CMS are employing different tagging working points, so 
combining their information allows us to constrain $\mu_c$. 
The typical tagging efficiencies are given in Table~\ref{table:tag}, and the combinations 
of working points in the analyses we use are given in Table~\ref{table:regions}. 
In the ATLAS~\cite{Aad:2014xzb} search there are two tagging points that have 
high and moderate rejection rates for $c$-jets, while CMS~\cite{Chatrchyan:2013zna} 
has four points with relatively high acceptance of $c$-jets.
Indeed, there are various values of $\epsilon^2_{c/b}$, categories (a)-(f) in 
Table~\ref{table:regions}. 
The tagging efficiencies do have a $p_{\rm T}^{\rm jet}$ dependence, but we have 
verified that the ratio of efficiencies, such as $\epsilon^2_{c/b}$, is less sensitive 
to the $p_{\rm T}^{\rm jet}$, see~\cite{ATLASbtag, Chatrchyan:2012jua}. 
Hereafter, we assume the efficiencies for each analysis to be constant.

\begin{table}[t!]
\begin{center}\begin{tabular}{|c|c|c||c|c|c|c|c|}
 \hline 
 ATLAS & Med & Tight  & CMS &Loose & Med1 & Med2 &Med3 \\\hline
 $\epsilon_b$ & 70\% & 50\% & $\epsilon_b$ & 88\% & 82\% & 78\% & 71\%\\
 $\epsilon_c$ & 20\% & 3.8\% & $\epsilon_c$ & 47\% & 34\% & 27\% & 21\%
 \\ \hline
 \end{tabular}
 \caption{The ATLAS and CMS $b$- and $c$-efficiencies for the different tagging criteria. 
          The CMS working points of CSV=0.244, 0.4, 0.5, and 0.677 are referred to as 
	  Loose, Med1, Med2, and Med3, respectively \cite{Chatrchyan:2012jua}. 
 \label{table:tag}}
\end{center}
%
\begin{center}\begin{tabular}{|l|c|c|c|c|c|c|c|}
 \hline 
& Figures & 1\textsuperscript{st} tag & 2\textsuperscript{nd} tag & $\epsilon^2_{c/b}$
\\ \hline
 (a)\,ATLAS & 11,12(a,b,d),13,17 & Med& Med & 0.082\\
 (b)\,ATLAS & 12(c) & Tight & Tight & 0.059\\ \hline
 (c)\,CMS & 10,11,12 &  Med1 &Med1 &0.18\\ 
  (d)\,CMS & 13 Left &  Med2 &Loose &0.19\\ 
  (e)\,CMS & 13 Right &  Med1 &Loose &0.23\\ 
   (f)\,CMS & 14 &  Med3 &Loose &0.16\\ 
  \hline
 \end{tabular} \caption{Summary of the experimental results used for the recast 
 of the $Vh(\bbbar)$ searches. Figures are taken from Refs.~\cite{Aad:2014xzb} and \cite{Chatrchyan:2013zna} 
 for ATLAS and CMS, respectively.\label{table:regions}}
\end{center}
\end{table}
\begin{figure}[t!]
\centering
\includegraphics[width=0.95\linewidth]{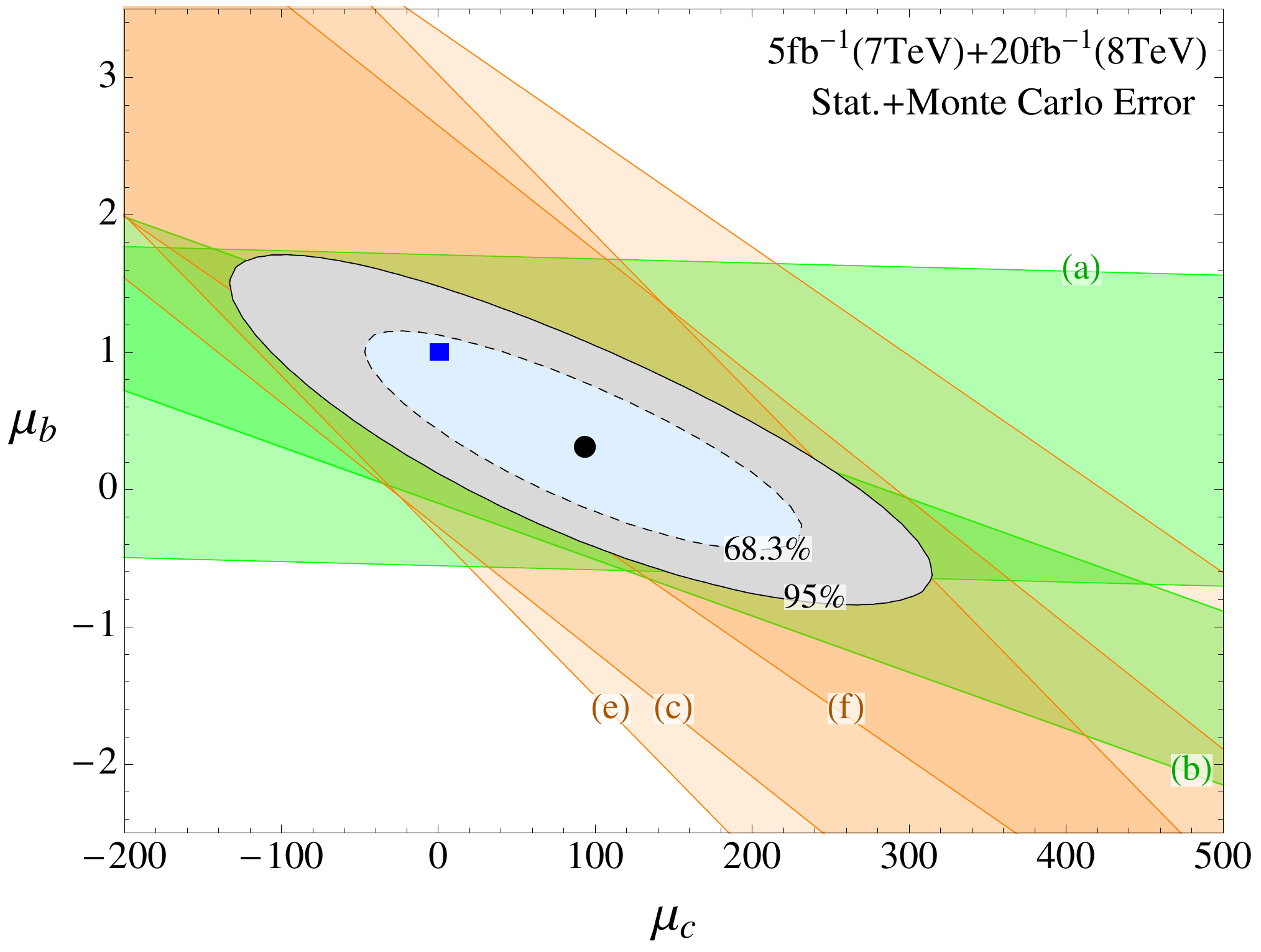}
\caption{68.3\%~CL (cyan) and 95\%~CL (gray) allowed regions in the 
$\mu_c$--$\mu_b$ plane.  
The best-fit (SM) point is indicated by the black circle (blue rectangle). 
The green(orange) bands are the 68.3\%~CL bands obtained from ATLAS(CMS) data. 
The labels (a)-(f) refer to the criteria in Table~\ref{table:regions}. 
Note that region~(d) is not shown because it is too broad.
\label{fig:mub-muc}
}
\end{figure}

For our recast study we proceed as follows. 
From existing data, summarized in Table~\ref{table:regions}, 
we use all the bins of the boosted decision tree output 
with $S/B\geq0.025$; those with lower ratios are simply 
background dominated. 
We then adopt the modified signal strength according to Eq.~\eqref{mubc} 
with $\epsilon^2_{c/b}$ depending on the category. 
We have constructed a likelihood function, $L(\mu_c,\mu_b)$, that is evaluated by a Poisson 
probability distribution convoluted with the Monte-Carlo systematic error with Gaussian weights.
For a parameter estimate, we use the likelihood ratio, 
\begin{equation}
	\lambda(\mu _c, \mu_b)=-2\log \frac{ L(\mu _c, \mu_b)}{ L(\hat{\mu}_c, \hat{\mu}_b)} \, ,
\end{equation}
where $\hat{\mu}_c$  and $\hat{\mu}_b$ are values at the best-fit point. 
In Fig.~\ref{fig:mub-muc}, we show the 68.3\%~CL and 95\%~CL contours as well as 
68.3\%~CL bands corresponding to each analysis (a)-(f). 
As discussed above, while the constraint of a given analysis is a flat direction
in the $\mu_c$--$\mu_b$ plane, the combination of different analyses disentangles 
the degeneracy leading to an ellipse. 
We further obtain the bound on $\mu_{c}$ with profiled $\mu_{b}$ 
(method of profile likelihood ratio \cite{Cowan:2010js}),
\begin{equation}
	\mu_c=95^{+90(175)}_{-95(180)}\  \rm at\ 68.3(95)\%\rm\ CL.\label{firstbound}
\end{equation}
This is the first direct and model-independent bound on the charm signal strength.

{\bf New production of $\boldsymbol{Vh}$ and charm Yukawa:} 
We would like to interpret the constraint of Eq.~\eqref{firstbound} as an 
upper bound on the charm Yukawa or, equivalently, on 
$\kappa_c \equiv y_c/y_c^{\rm SM}$. Similar $\kappa$ definitions hold for all 
Higgs couplings. 
Relative signs between $\kappa$'s do not affect our main 
results and we thus stick to $\kappa_X>0$. 

Assuming no modification of the production w.r.t.~the SM restricts the Higgs 
to charm signal strength to be
\begin{equation}
	\mu_c = \BR_\ccbar / {\BR^{\rm SM}_\ccbar}\lesssim 34\,.
	\label{noprod}
\end{equation}
The bound in Eq.~\eqref{firstbound} is weaker than the one in Eq.~\eqref{noprod}.
Thus, it cannot bound $\kappa_c$ from above, namely the inequality is satisfied 
even in the $\kappa_c\to \infty$ or $\BR_\ccbar\to1$ limit.

\begin{figure}[t!]
\centering
\includegraphics[width=0.55\linewidth]{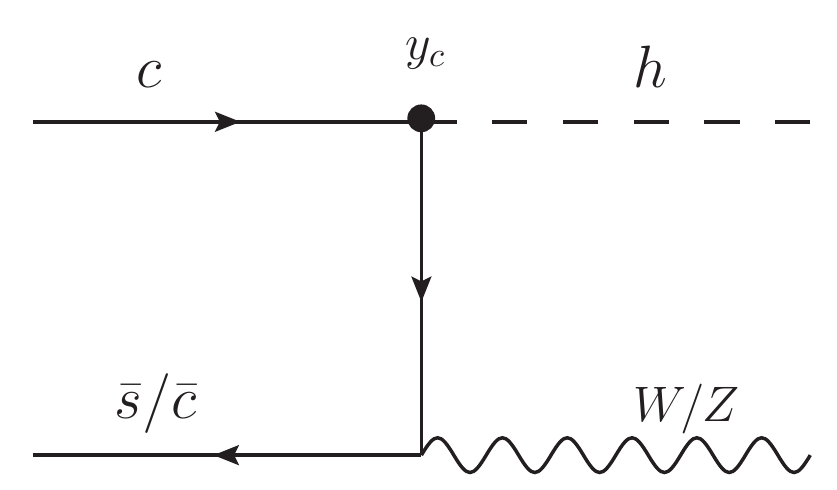}
\caption{Example diagram that modifies $Vh$ production when the charm-quark Yukawa 
is enhanced.\label{fig:Vhdiag}}
\end{figure}

\begin{figure*}[t]
\begin{center}
\includegraphics[width=0.95\linewidth]{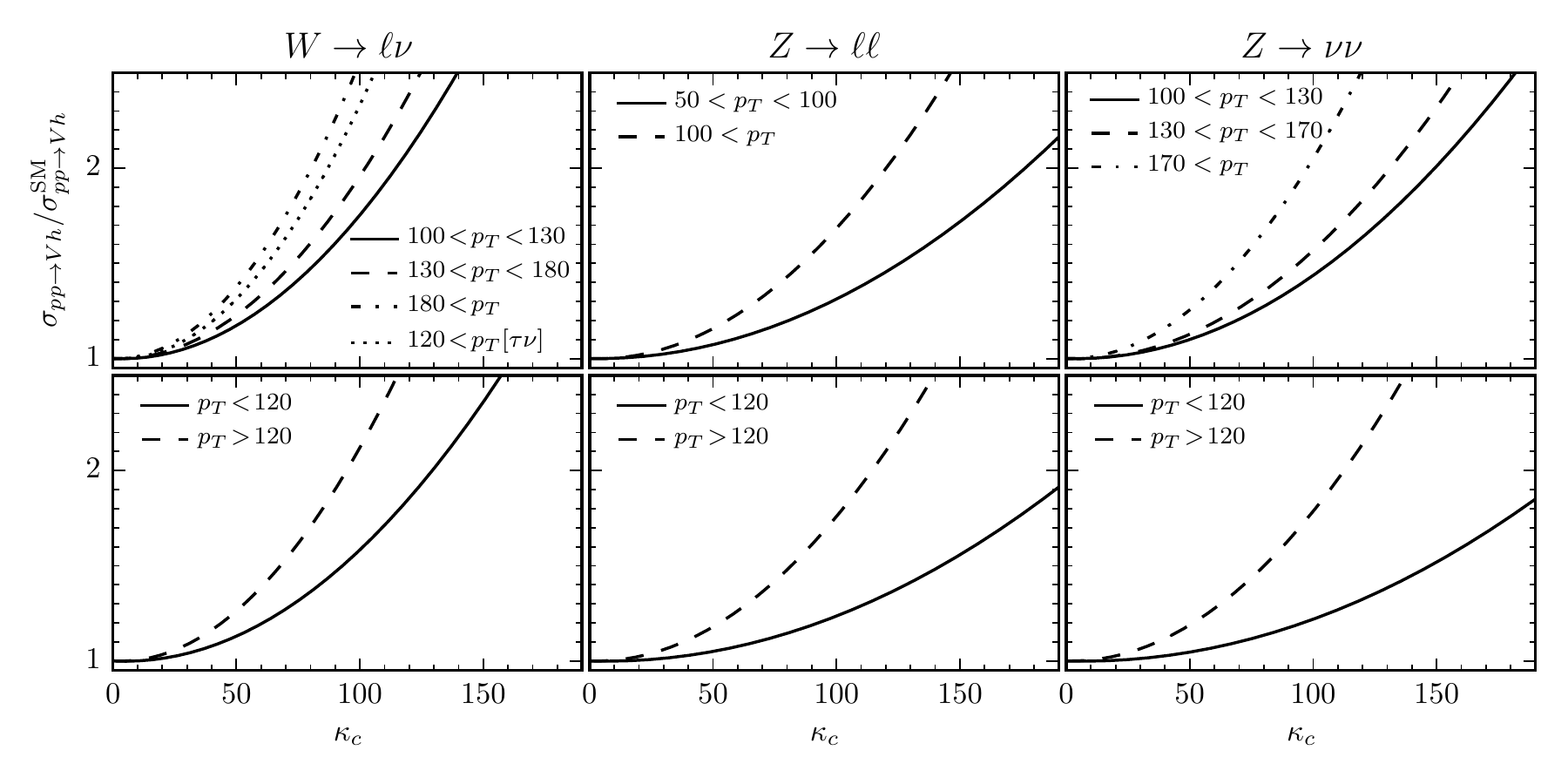}
\end{center}
\vspace{-20pt}
\caption{$Vh$ enhancement with $\kappa_c$ from the new production mechanism, using the preselection cuts of CMS and ATLAS. 
\label{fig:VHyc}}
\end{figure*}

However, as $\kappa_c$ (or more generally $\kappa_{u,d,s,c}$) becomes large, 
new contributions to the same final states, shown in Fig.~\ref{fig:Vhdiag}, 
become important and eliminate the ``runaway'' to arbitrarily large Yukawa. 
The contributions to the $Vh$ production cross section as a function of 
$\kappa_c$ are presented in Fig.~\ref{fig:VHyc} and roughly given by 
	\begin{equation}
	\frac{\sigma_{pp\to Vh}}{\sigma^{\rm SM}_{pp\to Vh}}\simeq1 +\left(\frac{\kappa_c}{\lambda_c}\right)^2\quad
	\text{with }
	\lambda_c=75{\rm -}200\, ,
	\end{equation}
for large $\kappa_c$, where the exact value of $\lambda_c$ depends on the channel. Here, the Higgs coupling to the $W/Z$ is assumed to be 
SM like, i.e. $\kappa_V=1$. 
We obtained these results using MadGraph 5.2~\cite{Alwall:2011uj} at the parton level and 
at leading order applying the CMS~\cite{Chatrchyan:2013zna} and ATLAS~\cite{Aad:2014xzb} selection cuts
for the LHC $8\,$TeV run. 
For a more complete treatment of the new production mechanisms, including the contributions 
from $u,d,s$ and also to final states with VBF-like topology, and comparison with
future machines we refer the reader to the companion paper~\cite{PSST}.

\begin{figure}[t!]
\centering
\includegraphics[width=0.95\linewidth]{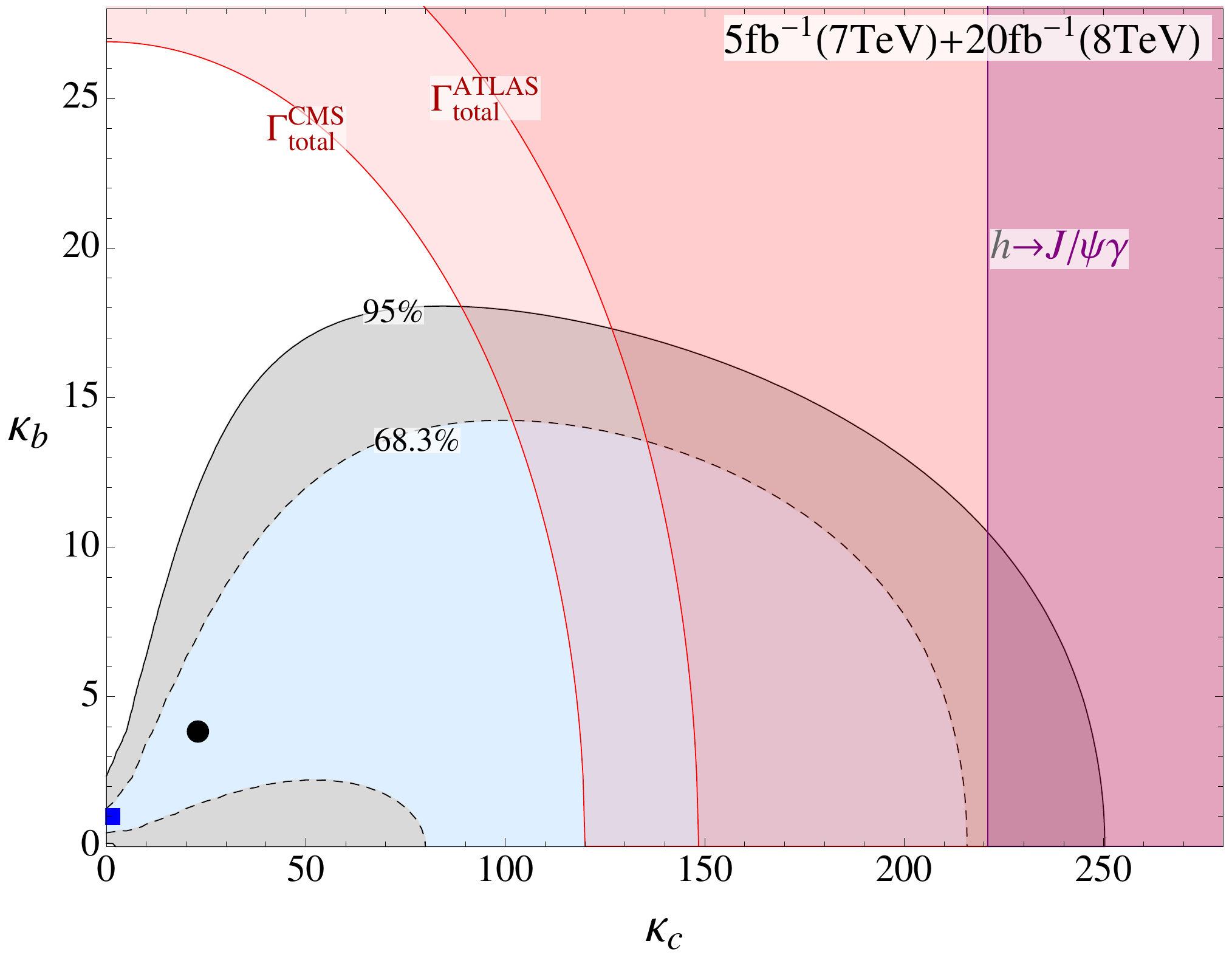}
\caption{68.3\%~CL (cyan) and 95\%~CL (gray) allowed regions of the 
recast study in the $\kappa_c$--$\kappa_b$ plane, with the best-fit (SM) point indicated
by the black circle(blue rectangle). 
Shaded areas represent the regions excluded by the total width  (ATLAS and CMS) and 
the exclusive Higgs decay of $h\to J/\psi \gamma$.
\label{fig:kappab-kappac}
}
\end{figure}

The new production mechanism significantly enhances the production cross section 
for large Yukawa, which is disfavoured by the $Vh$ data. 
In Fig.~\ref{fig:kappab-kappac} we thus combine ATLAS and CMS data to constrain
both $\kappa_c$ and $\kappa_b$. The allowed 68.3\,(95)\%~CL region is in blue\,(gray). The mapping between the signal strength and the Yukawa couplings, i.e. Fig.~\ref{fig:mub-muc} and Fig.~\ref{fig:kappab-kappac}, can be qualitatively understood by the relations  
\begin{equation}
	\mu_{c/b} \approx
	\left( 1 +\frac{\kappa_c^2}{\lambda_c^2} \right) 
	\frac{\kappa_{c/b}^2}{ 1+(\kappa^2_b-1)\BR^{\rm SM}_\bbbar + (\kappa^2_c-1)\BR^{\rm SM}_{\ccbar} } \, .
\end{equation}
From this also the mapping of the best fit points in the two plots can be understood. Profiling over $\kappa_b$ yields an upper bound on 
the charm Yukawa
\begin{equation}\label{inclusive}
\kappa_c\lesssim 234\,  \   \text{at\ 95\%\rm\ CL} \, .
\end{equation}
%

%

{\bf The total width:} 
Both ATLAS and CMS give a model independent bound on the Higgs total width 
from the invariant-mass distribution of the $h\to4\ell$ and $h\to\gamma\gamma$ signal. 
These bounds are limited by the experimental resolution of approximately $1\,$GeV. 
Assuming no interference with the background, the upper limits  
by ATLAS~\cite{Aad:2014aba} and CMS~\cite{Khachatryan:2014jba} are
	\begin{eqnarray}
	\!\!\!
	\Gamma_{\rm total} <\left\{
	\begin{array}{ll}
	2.4, \, 5.0 \, {\rm GeV}\ ({\rm CMS, \, ATLAS})\ \text{   }h\to\gamma\gamma &   \\
	3.4, \, 2.6 \, {\rm GeV}\ ({\rm CMS, \, ATLAS})\ \text{  }h\to4\ell&   
	\\	
	1.7 \, {\rm GeV}\ ({\rm CMS})\  \text{ combined $h \to \gamma\gamma, \, 4\ell$}
	\end{array} \right. \label{eq:GammaATLAS+CMS}
	\end{eqnarray}
at 95\% CL. 
This should be compared with the SM prediction of 
$\Gamma^{\rm SM}_{\rm total} = 4.07\,$MeV~\cite{Heinemeyer:2013tqa} for $m_h=125\,$GeV. 
We use the above upper bound on the total width to bound the charm Yukawa by assuming 
that the entire Higgs width is saturated by it
\begin{align}
	\kappa^2_c ~ \BR^{\rm SM}_{\ccbar} ~\Gamma_{\rm total}^{\rm SM} = 1.18 \times 10^{-4} \kappa^2_c \, {\rm GeV} < \Gamma_{\rm total}
\end{align}
with $\BR^{\rm SM}_{\ccbar} = 2.9\!\times\!10^{-2}\,$. 
The corresponding upper bounds at $95\%$ CL from Eq.~\eqref{eq:GammaATLAS+CMS} are
\begin{align}
	\kappa_c < 120 \,  (\text{CMS}),  \quad \label{width}
	\kappa_c < 150 \,  (\text{ATLAS}),
\end{align}
where in the case of ATLAS we have used the bound from $h\to4\ell$ and 
in the case of CMS the combined bound.

{\bf Interpretation of $\boldsymbol{h\to J/\psi \gamma}$:} \label{sec:HJpsi}
Very recently, ATLAS set the first bound on the exclusive Higgs 
decay to $J/\psi\gamma$~\cite{Aad:2015sda}
\begin{align} \label{eq:JpsiATLAS}
	\sigma \, \BR_{J/\psi \gamma} < 33\, {\rm fb}\ \   \text{at 95\% CL} \, .
\end{align}
Under the assumption of SM Higgs production, this can be interpreted as a bound 
of $\BR(h\to J/\psi\gamma)<1.5\!\times\!10^{-3}\,$. 
The partial width of $h\to J/\psi\gamma$ is given by~\cite{Bodwin:2014bpa}
\begin{equation}
\begin{split}
	\Gamma_{J/\psi \gamma}  
=	1.42 [&(1.0\pm0.017)\kappa_\gamma\\
&	- (0.087\pm0.012) \kappa_c ]^2 \times 10^{-8}\, {\rm GeV}\, .
\end{split}
\end{equation}
The dependence on the production mechanism and the Higgs total width can be canceled to a good approximation in the ratio between the bound (or measurement in the future) of the $h\to J/\psi\gamma$ rate and one of the other Higgs rate measurements with inclusive production, for example $h\to ZZ^*\to4\ell\,$. We define
\begin{equation}
 \label{eq:RJpsiZ}
\begin{split}
	\cR_{J/\psi,Z} 
=&	\frac{ \sigma \, \BR_{J/\psi \gamma} }{ \sigma \, \BR_{ZZ^*\to 4\ell}} 
\simeq	\frac{\Gamma_{J/\psi \gamma}  }{\Gamma_{ZZ^* \to 4\ell}}\\
=& \,2.79 \frac{ (\kappa_\gamma-0.087\kappa_c)^2 }{\kappa^2_V}\times 10^{-2}\, ,
\end{split}
\end{equation}
where a perfect cancellation of the production is assumed (correct to leading order) 
and $\BR^{\rm SM}_{ZZ^* \to 4\ell}=1.26\times 10^{-4}\,$~\cite{Heinemeyer:2013tqa}. 
Using Eq.~\eqref{eq:JpsiATLAS} and the $ZZ^*$ signal strength 
$\mu_{ZZ^*}=1.44^{+0.40}_{-0.33}$~\cite{Aad:2014eva} we extract 
\begin{align}
	\cR_{J/\psi,Z} = \frac{\sigma \, \BR_{J/\psi \gamma}}{\mu_{ZZ^*} \sigma_{\rm SM} \BR^{\rm SM}_{ZZ^* \to 4\ell} } < 9.3 \, ,
\end{align}
at 95\%~CL. Combining the last two equations leads to
\begin{align} \label{exclusive}
	-210\kappa_V + 11\kappa_\gamma < \kappa_c < 210\kappa_V + 11\kappa_\gamma \,.
\end{align}
This yields the bound $\kappa_c \lesssim 220$ 
assuming that $\kappa_{\gamma}$ and $\kappa_V$ (see discussion below) and 
also the Higgs decay width to a $Z$ and two leptons (\eg~$h\to Z\gamma^*\to 4\ell$) are all close to their 
respective SM values. 

{\bf Global analysis:} 
A global analysis of the Higgs data leads to an indirect bound on the 
Higgs total width and untagged decay  width, see \eg Refs.~\cite{Carmi:2012in,Giardino:2012dp,Espinosa:2012im,Ellis:2012hz,Falkowski:2013dza,Espinosa:2012vu,Dobrescu:2012td,Bechtle:2014ewa}. 
In the absence of non-SM production mechanisms, the allowed range for untagged 
decays is the leading bound on the charm Yukawa.
For this, we can safely ignore non-SM $Vh$ and VBF-like production enhancements 
because they are found to be negligible for $\kappa_c \lesssim 50$. 
The allowed range of $\kappa_V$ from EW precision data assuming a cutoff scale of 
$3\,$TeV is $\kappa_V = 1.08\pm0.07\,$~\cite{Falkowski:2013dza}. 
This, along with the Higgs measurement of VBF and gluon fusion in $WW^*$, $ZZ^*$, 
and $\tau\bar\tau$ final states, results in a much stronger bound on the total 
Higgs width than the direct measurement. 

Following the analysis of Ref.~\cite{Delaunay:2013pja}, we consider the 
current available Higgs data from ATLAS~\cite{ATLAStth,Aad:2014eva,Aad:2014eha,Aad:2014xzb,Aad:2015vsa,TheATLAScollaboration:2013hia,ATLAS060,ATLAS061}, CMS~\cite{Chatrchyan:2013iaa,Khachatryan:2014jba,Chatrchyan:2013mxa,Khachatryan:2014ira,Khachatryan:2014aep,Chatrchyan:2014nva,CMS:2013xda,Chatrchyan:2013zna,Khachatryan:2014qaa} and Tevatron~\cite{Aaltonen:2013ipa,Abazov:2013gmz}, extracted by using Ref.~\cite{Bechtle:2013xfa}, along with the EW data as 
in Ref.~\cite{Falkowski:2013dza}. We find that the 95\%~CL allowed range for the charm Yukawa is
\beq
\kappa_c \lesssim 6.2\,  ,  \label{global} 
\eeq
where all the Higgs couplings (including $h\to WW,\, ZZ,\, \gamma\gamma,\, gg,\, Z\gamma,\, b\bar{b},\, \tau\bar\tau$) 
were allowed to vary from their SM values.  Allowing the up-quark Yukawa also to vary does not change this bound. Note that the bound in Eq.~\eqref{global} depends on the global fit assumption, in particular the LEP constraints, and as such carries model dependence. 

The ratio between the on-shell and the off-shell $h\to ZZ^{(*)}$ rates can probe the Higgs width~\cite{Caola:2013yja}. The current bounds are at the order of $\Gamma_{\rm total}/\Gamma^{\rm SM}_{\rm total}\lesssim 5.4\, , 7.7 $~ from CMS~\cite{Khachatryan:2014iha} and ATLAS~\cite{ATLASoffshell}, respectively.
This corresponds to $\kappa_c\lesssim14\, , 16$. However, as pointed out in Ref.~\cite{Englert:2014aca} these bounds are model dependent. Thus, we do not further consider this bound in our analysis. 
We mention, that also low-energy processes can indirectly constrain 
light-quark Yukawas, see for example Refs.~\cite{Isidori:2013cla, Harnik:2012pb,Goertz:2014qia}.   

{\bf Higgs--quark non-universality: }
We now turn to provide a lower bound on the top Yukawa coupling in order to 
compare it with the upper bounds on the charm Yukawa coupling obtained above.
A comparison with $t\bar{t}h$ data allows us to show that 
current data eliminates the possibility that the Higgs couples to quarks
in a universal way, as is expected in the SM.
As mentioned in Eq.~\eqref{eq:3rdavg}, a naive average of the ATLAS and 
CMS results yields $\mu_{t\bar{t}h}=2.4\pm0.8$. 
This leads to a lower bound on the top Yukawa (at 95\% CL), 
\begin{align}
	\kappa_t > 0.9\sqrt{ \frac{ \BR^{\rm SM}_{\rm finals} }{ \BR_{\rm finals}} } >0.9  \label{kappat}\, ,
\end{align}
where $\BR_{\rm finals}$ stands for the final states that were considered 
by the collaborations in the $t\bar{t}h$ measurements. 
The last inequality is valid in case that the Higgs to charm pairs is 
the dominant partial width (as is expected in the case where our rather weak bounds 
obtained above are saturated). 
In the special case where the dominant decays are to charms 
and $\tau$'s, namely $\kappa_\tau\gg1$,
we have $\mu_{{\rm VBF},\tau}>2$, which is excluded by data~\cite{Aad:2015vsa,Chatrchyan:2014nva}. 
We thus conclude that 
\begin{align}
	 \frac{y_c }{y_t} =\frac{\kappa_c}{\kappa_t} \frac{y^{\rm SM}_c}{y^{\rm SM}_t}\simeq {1\over 280} \times \frac{\kappa_c}{\kappa_t} 
	\quad \Rightarrow\quad
	y_c< y_t \, ,
\end{align}
where the last inequality is based on comparison of Eqs.~\eqref{inclusive},~\eqref{width},~\eqref{exclusive} 
and~\eqref{global} with Eq.~\eqref{kappat}. 
We therefore conclude that the Yukawa couplings of the up-type quarks are non-universal. 

{\bf Summary of LHC constraints:}
In Fig.~\ref{fig:kappab-kappac} we present bounds on Higgs couplings
from the $Vh$ recast, the total width measurements, 
and the exclusive decay to $J/\psi \gamma$, on the $\kappa_c$--$\kappa_b$ plane. 
We see that the relatively robust bounds from the $Vh$ recast and the total width 
measurements are  of same order of magnitude and also complement each other. 

\begin{figure}[t!]
\centering
\includegraphics[width=1.0\linewidth]{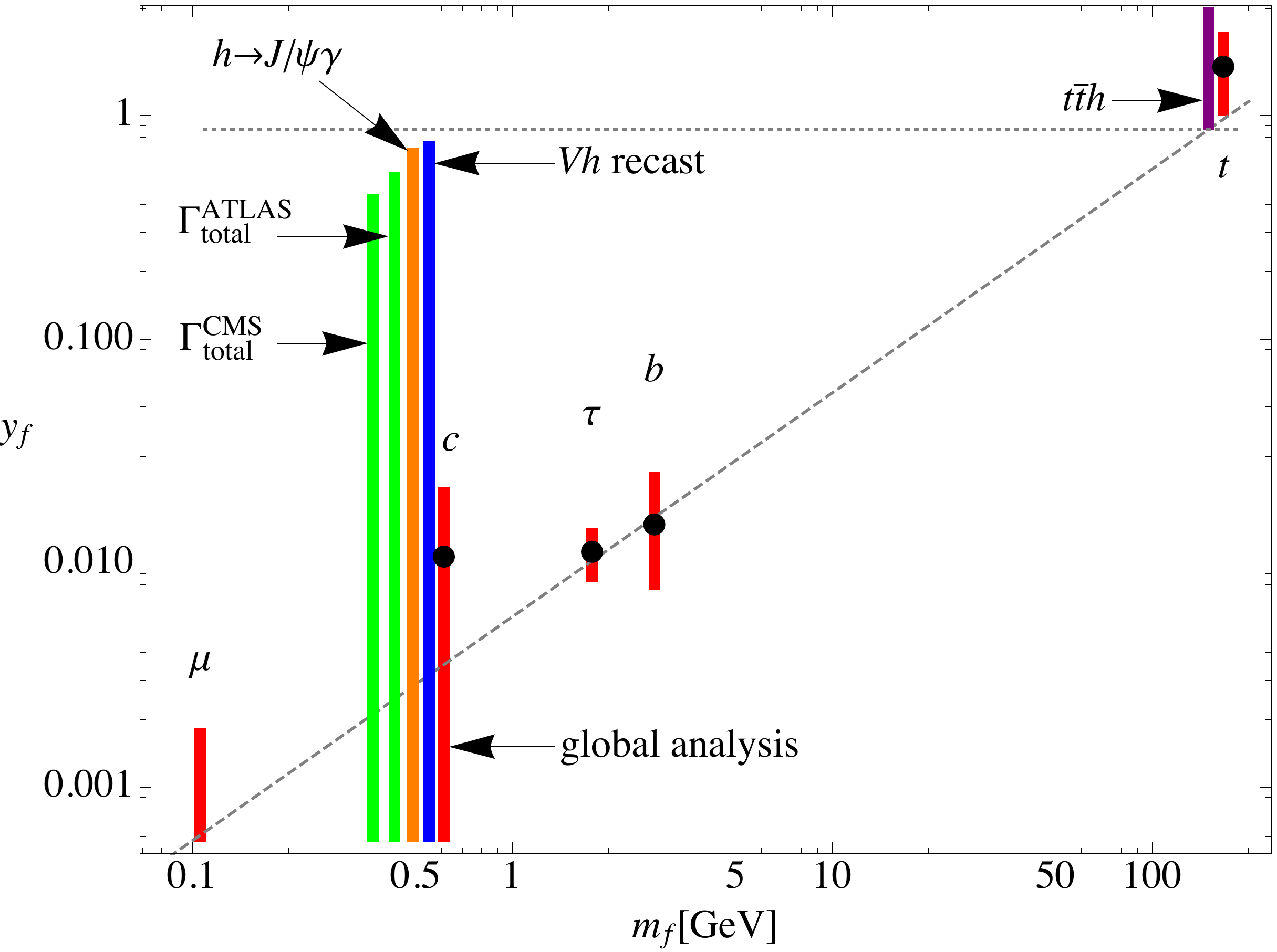}
\caption{Summary of current constraints on the Higgs couplings to fermions including 
the new bounds on the charm Yukawa. 
\label{fig:money}}
\end{figure}

In Fig.~\ref{fig:money} we show the  95\% CL regions for the Higgs couplings to 
fermions as a function of their masses based on the global analysis and  we have added
the bounds obtained above regarding the charm Yukawa coupling. 

An improvement of the bound on the charm signal strength can be achieved 
by adopting the charm-tagging~\cite{ATL-PHYS-PUB-2015-001}. 
We estimate the sensitivity from current data as follows. 
We rescale the expected number of signal and background events of the  8\,TeV ATLAS analysis 
(Table 8 of Ref.~\cite{Aad:2014xzb}) according to the efficiencies of the charm-tagging \cite{Aad:2015gna}, 
	\begin{eqnarray}
	\epsilon_b=13\%\,, \ \  \epsilon_c=19\%\,, \ \ \epsilon_l=0.5\%\,, 
	\end{eqnarray}
 where $\epsilon_l$ is efficiency to tag light jets. 
Here, we assume that medium $b$-tagging in Table~\ref{table:tag} ($\epsilon_l=1.25\%$) 
is used in the analysis and that the decomposition of $W(Z)+$heavy-flavor quarks 
background is 35(20)\% $W(Z)+c\bar{c}$ and 65(80)\% $W(Z)+b\bar{b}$. 
We combine the rescaled ATLAS analysis with the CMS results (c)-(f) in 
Table~\ref{table:regions} and  obtain an uncertainty of
 \begin{align}
 	\Delta\mu_c \simeq 50\, (107)\, ,
 \end{align}
at 68.3 (95)\% CL. We see that even with the same luminosity the error is significantly reduced 
with respect to the one in Eq.~\eqref{firstbound}. 
 
{\bf Future LHC prospects:}
Finally, we estimate the future sensitivity at the LHC.  
We utilize results of Tables~6-9 in Ref.~\cite{ATL-PHYS-PUB-2014-011} where ATLAS 
performed a dedicated Monte Carlo study of $Vh(b\bar{b})$ in the 1- and 2-lepton final states 
for LHC~run~II with 300 fb$^{-1}$ and  LHC high-luminosity upgrade (HL-LHC) with 3000 fb$^{-1}$\,at 14 TeV. 
From the given working point of medium $b$-tagging, we rescale the signal and background of 
1-lepton final state to those in charm-tagging. 
We leave the 2-lepton analysis as original because, as discussed, we need at least two 
working points to extract $\mu_b$ and $\mu_c$ independently. We then also assume that
the same analysis can be performed by CMS.

\begin{figure}[t!]
\centering
\includegraphics[width=0.9\linewidth]{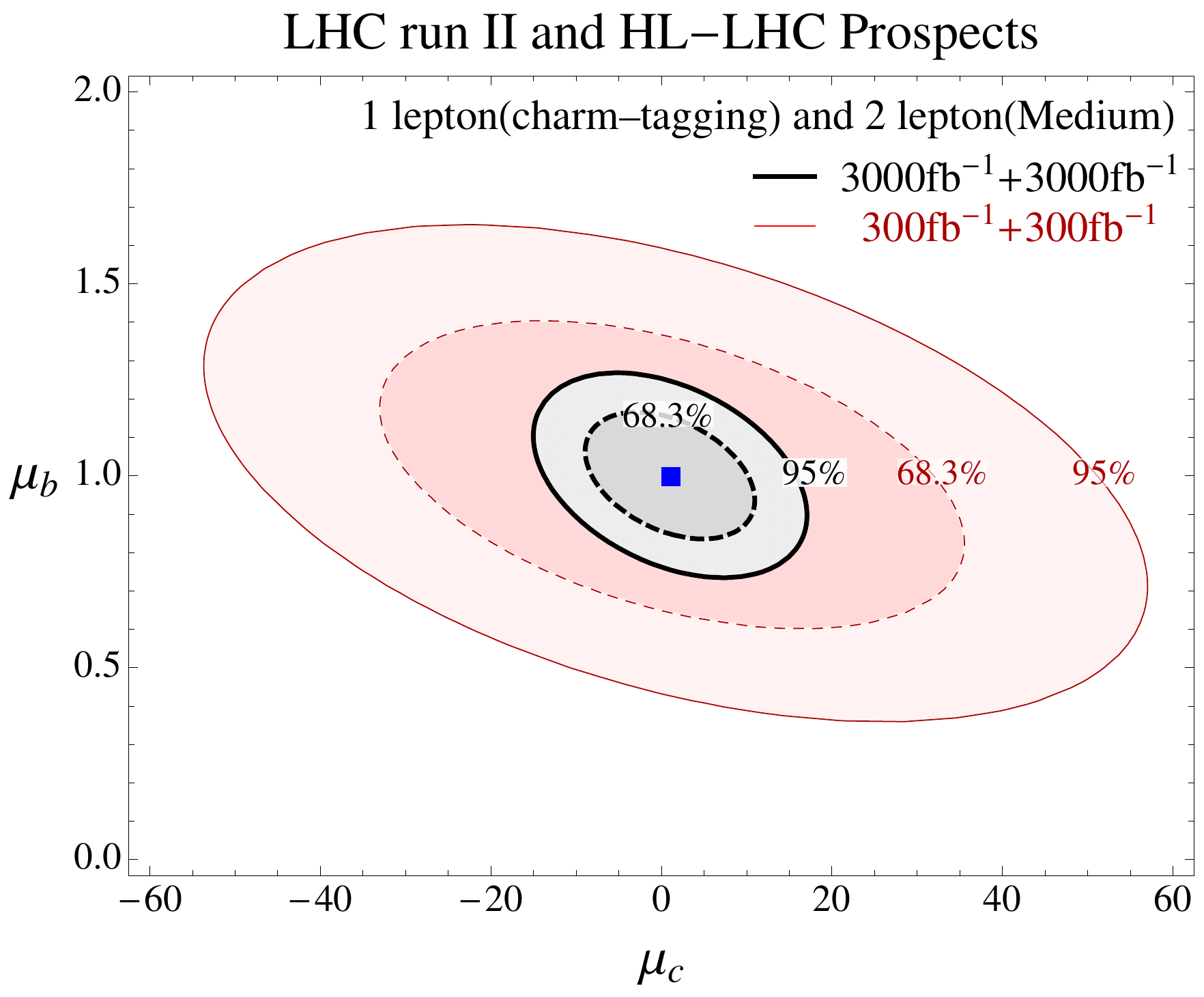}
\caption{Expected reach for the signal-strength measurement of $h\to b\bar b$ 
and $h\to c\bar c$ at LHC run II and HL-LHC: The black-thick (purple-thin) curves 
correspond to the reach with 3000 (300) fb$^{-1}$. 
The solid (dashed) ones correspond to 68.3\,(95) \% CL. 
The SM expectation is $\mu_{b,c}=1\,.$ \label{fig:mub-mucFF}
}
\end{figure}

The future sensitivity reach for $\mu_c$ is shown as ellipses in 
the $\mu_c$--$\mu_b$ plane in Fig.~\ref{fig:mub-mucFF}. 
Here, we take into account only the statistical error. 
The expected uncertainty with profiled $\mu_b$ reads 
	\begin{equation}
\begin{split}
	\Delta\mu_c=\left\{
	\begin{array}{ccc}
	23\,(45)&\text{with $2\times300$ fb$^{-1}$}  \\
	6.5\,(13) & \text{with $2\times3000$ fb$^{-1}$}   \\
	\end{array} \right.
\end{split}
	\end{equation}
at  68.3\,(95)\% CL. Compared to the result of LHC~run~I, 
the uncertainty is improved by roughly an order of magnitude 
with 3000 fb$^{-1}$ thanks to charm-tagging. 
In the future, one may hope that the charm-tagging performance 
will be further optimized. 
As an example for such a case, we have considered the following 
improved  charm-tagging point $\epsilon_b=20\,$\%, $\epsilon_c=40\,$\% 
and $\epsilon_l=1.25\,$\%. 
As a consequence the bounds will be further strengthened,  
$\Delta\mu_c\simeq 20\,(6.5) $ at 95\,\%~CL with integrated luminosity 
of $2\times 300\,(2\times3000)\,$fb$^{-1}$.

{\bf  Conclusions:} 
We have performed four different analyses to constrain the charm Yukawa 
and obtained the following bounds
	\begin{eqnarray}
	{y_c\over y_c^{\rm SM}} \lesssim 234, \ 120 \,(140),\  220,\  6.2, 
	\end{eqnarray}
that correspond to: a recast of the $h\to b\bar b$ searches, 
the direct bound on the Higgs total width at CMS (ATLAS), 
the exclusive decay of $h\to J/\psi \gamma $, 
and the global analysis, respectively. 
Together with the $t\bar{t}h$ analyses of ATLAS and CMS we  
conclude that the Higgs coupling to the top and charm quarks 
is not universal. 
We further point out two new production mechanisms, related to $Vh$  and VBF processes
that become important when the first two generation quarks have enhanced couplings to 
the Higgs. 
In conjunction with a future measurement at an electron-positron collider (linear or circular) 
the former mechanism is sensitive to the Higgs--light-quark couplings.  
We also provide projections for the sensitivity of the LHC experiments 
to the charm Yukawa by adopting a dedicated charm-tagging analysis resulting in an order of magnitude improvement. 
Finally, we point out that with the recent installation of the 
Insertable B-Layer~(IBL) sub-detector~\cite{Capeans:1291633}, 
the ATLAS capability for charm-tagging is expected to be further improved 
enhancing the sensitivity to the Higgs--charm coupling. 

{\it\bf Acknowledgments:} 
We thank Anadi~Canepa, Eilam~Gross, Fabio~Maltoni, Frank~Petriello and 
Tim~Stefaniak for helpful discussions. We also acknowledge help from 
the ATLAS collaboration for providing us with details of the analysis 
of Ref.~\cite{ATL-PHYS-PUB-2014-011}.
The work of KT is supported in part by the
Grant-in-Aid for JSPS Fellows. The work of GP is supported by the ERC, IRG and ISF grants. 

\bibliographystyle{aipnum4-1}
\bibliographystyle{apsrev4-1}
\bibliography{ref}

\begin{thebibliography}{73}%
\makeatletter
\providecommand \@ifxundefined [1]{%
 \@ifx{#1\undefined}
}%
\providecommand \@ifnum [1]{%
 \ifnum #1\expandafter \@firstoftwo
 \else \expandafter \@secondoftwo
 \fi
}%
\providecommand \@ifx [1]{%
 \ifx #1\expandafter \@firstoftwo
 \else \expandafter \@secondoftwo
 \fi
}%
\providecommand \natexlab [1]{#1}%
\providecommand \enquote  [1]{``#1''}%
\providecommand \bibnamefont  [1]{#1}%
\providecommand \bibfnamefont [1]{#1}%
\providecommand \citenamefont [1]{#1}%
\providecommand \href@noop [0]{\@secondoftwo}%
\providecommand \href [0]{\begingroup \@sanitize@url \@href}%
\providecommand \@href[1]{\@@startlink{#1}\@@href}%
\providecommand \@@href[1]{\endgroup#1\@@endlink}%
\providecommand \@sanitize@url [0]{\catcode `\\12\catcode `\$12\catcode
  `\&12\catcode `\#12\catcode `\^12\catcode `\_12\catcode `\%12\relax}%
\providecommand \@@startlink[1]{}%
\providecommand \@@endlink[0]{}%
\providecommand \url  [0]{\begingroup\@sanitize@url \@url }%
\providecommand \@url [1]{\endgroup\@href {#1}{\urlprefix }}%
\providecommand \urlprefix  [0]{URL }%
\providecommand \Eprint [0]{\href }%
\providecommand \doibase [0]{http://dx.doi.org/}%
\providecommand \selectlanguage [0]{\@gobble}%
\providecommand \bibinfo  [0]{\@secondoftwo}%
\providecommand \bibfield  [0]{\@secondoftwo}%
\providecommand \translation [1]{[#1]}%
\providecommand \BibitemOpen [0]{}%
\providecommand \bibitemStop [0]{}%
\providecommand \bibitemNoStop [0]{.\EOS\space}%
\providecommand \EOS [0]{\spacefactor3000\relax}%
\providecommand \BibitemShut  [1]{\csname bibitem#1\endcsname}%
\let\auto@bib@innerbib\@empty
\bibitem [{\citenamefont {Aad}\ \emph {et~al.}(2012)\citenamefont {Aad} \emph
  {et~al.}}]{Aad:2012tfa}%
  \BibitemOpen
  \bibfield  {author} {\bibinfo {author} {\bibfnamefont {G.}~\bibnamefont
  {Aad}} \emph {et~al.} (\bibinfo {collaboration} {ATLAS Collaboration}),\
  }\href {\doibase 10.1016/j.physletb.2012.08.020} {\bibfield  {journal}
  {\bibinfo  {journal} {Phys.Lett.}\ }\textbf {\bibinfo {volume} {B716}},\
  \bibinfo {pages} {1} (\bibinfo {year} {2012})},\ \Eprint
  {http://arxiv.org/abs/1207.7214} {arXiv:1207.7214 [hep-ex]} \BibitemShut
  {NoStop}%
\bibitem [{\citenamefont {Chatrchyan}\ \emph {et~al.}(2012)\citenamefont
  {Chatrchyan} \emph {et~al.}}]{Chatrchyan:2012ufa}%
  \BibitemOpen
  \bibfield  {author} {\bibinfo {author} {\bibfnamefont {S.}~\bibnamefont
  {Chatrchyan}} \emph {et~al.} (\bibinfo {collaboration} {CMS Collaboration}),\
  }\href {\doibase 10.1016/j.physletb.2012.08.021} {\bibfield  {journal}
  {\bibinfo  {journal} {Phys.Lett.}\ }\textbf {\bibinfo {volume} {B716}},\
  \bibinfo {pages} {30} (\bibinfo {year} {2012})},\ \Eprint
  {http://arxiv.org/abs/1207.7235} {arXiv:1207.7235 [hep-ex]} \BibitemShut
  {NoStop}%
\bibitem [{The ATLAS Collaboration, ATLAS-CONF-2014-011,
  ATLAS-COM-CONF-2014-004()}]{ATLAStth}%
  \BibitemOpen
  \bibfield  {author} {The ATLAS Collaboration, ATLAS-CONF-2014-011,
  ATLAS-COM-CONF-2014-004,\ }\href@noop {} {\  (\bibinfo {year}
  {2014})}\BibitemShut {NoStop}%
\bibitem [{\citenamefont {Aad}\ \emph {et~al.}(2014{\natexlab{a}})\citenamefont
  {Aad} \emph {et~al.}}]{Aad:2014xzb}%
  \BibitemOpen
  \bibfield  {author} {\bibinfo {author} {\bibfnamefont {G.}~\bibnamefont
  {Aad}} \emph {et~al.} (\bibinfo {collaboration} {ATLAS Collaboration}),\
  }\href@noop {} {\  (\bibinfo {year} {2014}{\natexlab{a}})},\ \Eprint
  {http://arxiv.org/abs/1409.6212} {arXiv:1409.6212 [hep-ex]} \BibitemShut
  {NoStop}%
\bibitem [{\citenamefont {Aad}\ \emph {et~al.}(2015{\natexlab{a}})\citenamefont
  {Aad} \emph {et~al.}}]{Aad:2015vsa}%
  \BibitemOpen
  \bibfield  {author} {\bibinfo {author} {\bibfnamefont {G.}~\bibnamefont
  {Aad}} \emph {et~al.} (\bibinfo {collaboration} {ATLAS Collaboration}),\
  }\href@noop {} {\  (\bibinfo {year} {2015}{\natexlab{a}})},\ \Eprint
  {http://arxiv.org/abs/1501.04943} {arXiv:1501.04943 [hep-ex]} \BibitemShut
  {NoStop}%
\bibitem [{\citenamefont {Khachatryan}\ \emph
  {et~al.}(2014{\natexlab{a}})\citenamefont {Khachatryan} \emph
  {et~al.}}]{Khachatryan:2014qaa}%
  \BibitemOpen
  \bibfield  {author} {\bibinfo {author} {\bibfnamefont {V.}~\bibnamefont
  {Khachatryan}} \emph {et~al.} (\bibinfo {collaboration} {CMS
  Collaboration}),\ }\href {\doibase 10.1007/JHEP09(2014)087,
  10.1007/JHEP10(2014)106} {\bibfield  {journal} {\bibinfo  {journal} {JHEP}\
  }\textbf {\bibinfo {volume} {1409}},\ \bibinfo {pages} {087} (\bibinfo {year}
  {2014}{\natexlab{a}})},\ \Eprint {http://arxiv.org/abs/1408.1682}
  {arXiv:1408.1682 [hep-ex]} \BibitemShut {NoStop}%
\bibitem [{\citenamefont {Chatrchyan}\ \emph
  {et~al.}(2014{\natexlab{a}})\citenamefont {Chatrchyan} \emph
  {et~al.}}]{Chatrchyan:2013zna}%
  \BibitemOpen
  \bibfield  {author} {\bibinfo {author} {\bibfnamefont {S.}~\bibnamefont
  {Chatrchyan}} \emph {et~al.} (\bibinfo {collaboration} {CMS Collaboration}),\
  }\href {\doibase 10.1103/PhysRevD.89.012003} {\bibfield  {journal} {\bibinfo
  {journal} {Phys.Rev.}\ }\textbf {\bibinfo {volume} {D89}},\ \bibinfo {pages}
  {012003} (\bibinfo {year} {2014}{\natexlab{a}})},\ \Eprint
  {http://arxiv.org/abs/1310.3687} {arXiv:1310.3687 [hep-ex]} \BibitemShut
  {NoStop}%
\bibitem [{\citenamefont {Chatrchyan}\ \emph
  {et~al.}(2014{\natexlab{b}})\citenamefont {Chatrchyan} \emph
  {et~al.}}]{Chatrchyan:2014nva}%
  \BibitemOpen
  \bibfield  {author} {\bibinfo {author} {\bibfnamefont {S.}~\bibnamefont
  {Chatrchyan}} \emph {et~al.} (\bibinfo {collaboration} {CMS Collaboration}),\
  }\href {\doibase 10.1007/JHEP05(2014)104} {\bibfield  {journal} {\bibinfo
  {journal} {JHEP}\ }\textbf {\bibinfo {volume} {1405}},\ \bibinfo {pages}
  {104} (\bibinfo {year} {2014}{\natexlab{b}})},\ \Eprint
  {http://arxiv.org/abs/1401.5041} {arXiv:1401.5041 [hep-ex]} \BibitemShut
  {NoStop}%
\bibitem [{\citenamefont {Aad}\ \emph {et~al.}(2014{\natexlab{b}})\citenamefont
  {Aad} \emph {et~al.}}]{Aad:2014xva}%
  \BibitemOpen
  \bibfield  {author} {\bibinfo {author} {\bibfnamefont {G.}~\bibnamefont
  {Aad}} \emph {et~al.} (\bibinfo {collaboration} {ATLAS Collaboration}),\
  }\href {\doibase 10.1016/j.physletb.2014.09.008} {\bibfield  {journal}
  {\bibinfo  {journal} {Phys.Lett.}\ }\textbf {\bibinfo {volume} {B738}},\
  \bibinfo {pages} {68} (\bibinfo {year} {2014}{\natexlab{b}})},\ \Eprint
  {http://arxiv.org/abs/1406.7663} {arXiv:1406.7663 [hep-ex]} \BibitemShut
  {NoStop}%
\bibitem [{\citenamefont {Khachatryan}\ \emph
  {et~al.}(2014{\natexlab{b}})\citenamefont {Khachatryan} \emph
  {et~al.}}]{Khachatryan:2014aep}%
  \BibitemOpen
  \bibfield  {author} {\bibinfo {author} {\bibfnamefont {V.}~\bibnamefont
  {Khachatryan}} \emph {et~al.} (\bibinfo {collaboration} {CMS
  Collaboration}),\ }\href@noop {} {\  (\bibinfo {year}
  {2014}{\natexlab{b}})},\ \Eprint {http://arxiv.org/abs/1410.6679}
  {arXiv:1410.6679 [hep-ex]} \BibitemShut {NoStop}%
\bibitem [{\citenamefont {Chivukula}\ \emph {et~al.}(2007)\citenamefont
  {Chivukula}, \citenamefont {Christensen}, \citenamefont {Coleppa},\ and\
  \citenamefont {Simmons}}]{Chivukula:2007mw}%
  \BibitemOpen
  \bibfield  {author} {\bibinfo {author} {\bibfnamefont {R.~S.}\ \bibnamefont
  {Chivukula}}, \bibinfo {author} {\bibfnamefont {N.~D.}\ \bibnamefont
  {Christensen}}, \bibinfo {author} {\bibfnamefont {B.}~\bibnamefont
  {Coleppa}}, \ and\ \bibinfo {author} {\bibfnamefont {E.~H.}\ \bibnamefont
  {Simmons}},\ }\href {\doibase 10.1103/PhysRevD.75.073018} {\bibfield
  {journal} {\bibinfo  {journal} {Phys.Rev.}\ }\textbf {\bibinfo {volume}
  {D75}},\ \bibinfo {pages} {073018} (\bibinfo {year} {2007})},\ \Eprint
  {http://arxiv.org/abs/hep-ph/0702281} {arXiv:hep-ph/0702281 [HEP-PH]}
  \BibitemShut {NoStop}%
\bibitem [{\citenamefont {Marciano}\ \emph {et~al.}(1989)\citenamefont
  {Marciano}, \citenamefont {Valencia},\ and\ \citenamefont
  {Willenbrock}}]{Marciano:1989ns}%
  \BibitemOpen
  \bibfield  {author} {\bibinfo {author} {\bibfnamefont {W.~J.}\ \bibnamefont
  {Marciano}}, \bibinfo {author} {\bibfnamefont {G.}~\bibnamefont {Valencia}},
  \ and\ \bibinfo {author} {\bibfnamefont {S.}~\bibnamefont {Willenbrock}},\
  }\href {\doibase 10.1103/PhysRevD.40.1725} {\bibfield  {journal} {\bibinfo
  {journal} {Phys.Rev.}\ }\textbf {\bibinfo {volume} {D40}},\ \bibinfo {pages}
  {1725} (\bibinfo {year} {1989})}\BibitemShut {NoStop}%
\bibitem [{\citenamefont {Appelquist}\ and\ \citenamefont
  {Chanowitz}(1987)}]{Appelquist:1987cf}%
  \BibitemOpen
  \bibfield  {author} {\bibinfo {author} {\bibfnamefont {T.}~\bibnamefont
  {Appelquist}}\ and\ \bibinfo {author} {\bibfnamefont {M.~S.}\ \bibnamefont
  {Chanowitz}},\ }\href {\doibase 10.1103/PhysRevLett.59.2405} {\bibfield
  {journal} {\bibinfo  {journal} {Phys.Rev.Lett.}\ }\textbf {\bibinfo {volume}
  {59}},\ \bibinfo {pages} {2405} (\bibinfo {year} {1987})}\BibitemShut
  {NoStop}%
\bibitem [{\citenamefont {Maltoni}\ \emph {et~al.}(2001)\citenamefont
  {Maltoni}, \citenamefont {Niczyporuk},\ and\ \citenamefont
  {Willenbrock}}]{Maltoni:2000iq}%
  \BibitemOpen
  \bibfield  {author} {\bibinfo {author} {\bibfnamefont {F.}~\bibnamefont
  {Maltoni}}, \bibinfo {author} {\bibfnamefont {J.~M.}\ \bibnamefont
  {Niczyporuk}}, \ and\ \bibinfo {author} {\bibfnamefont {S.}~\bibnamefont
  {Willenbrock}},\ }\href {\doibase 10.1103/PhysRevLett.86.212} {\bibfield
  {journal} {\bibinfo  {journal} {Phys.Rev.Lett.}\ }\textbf {\bibinfo {volume}
  {86}},\ \bibinfo {pages} {212} (\bibinfo {year} {2001})},\ \Eprint
  {http://arxiv.org/abs/hep-ph/0006358} {arXiv:hep-ph/0006358 [hep-ph]}
  \BibitemShut {NoStop}%
\bibitem [{\citenamefont {Dicus}\ and\ \citenamefont
  {He}(2005)}]{Dicus:2004rg}%
  \BibitemOpen
  \bibfield  {author} {\bibinfo {author} {\bibfnamefont {D.~A.}\ \bibnamefont
  {Dicus}}\ and\ \bibinfo {author} {\bibfnamefont {H.-J.}\ \bibnamefont {He}},\
  }\href {\doibase 10.1103/PhysRevD.71.093009} {\bibfield  {journal} {\bibinfo
  {journal} {Phys.Rev.}\ }\textbf {\bibinfo {volume} {D71}},\ \bibinfo {pages}
  {093009} (\bibinfo {year} {2005})},\ \Eprint
  {http://arxiv.org/abs/hep-ph/0409131} {arXiv:hep-ph/0409131 [hep-ph]}
  \BibitemShut {NoStop}%
\bibitem [{\citenamefont {Delaunay}\ \emph {et~al.}(2013)\citenamefont
  {Delaunay}, \citenamefont {Grojean},\ and\ \citenamefont
  {Perez}}]{Delaunay:2013iia}%
  \BibitemOpen
  \bibfield  {author} {\bibinfo {author} {\bibfnamefont {C.}~\bibnamefont
  {Delaunay}}, \bibinfo {author} {\bibfnamefont {C.}~\bibnamefont {Grojean}}, \
  and\ \bibinfo {author} {\bibfnamefont {G.}~\bibnamefont {Perez}},\ }\href
  {\doibase 10.1007/JHEP09(2013)090} {\bibfield  {journal} {\bibinfo  {journal}
  {JHEP}\ }\textbf {\bibinfo {volume} {1309}},\ \bibinfo {pages} {090}
  (\bibinfo {year} {2013})},\ \Eprint {http://arxiv.org/abs/1303.5701}
  {arXiv:1303.5701 [hep-ph]} \BibitemShut {NoStop}%
\bibitem [{\citenamefont {Delaunay}\ \emph
  {et~al.}(2014{\natexlab{a}})\citenamefont {Delaunay}, \citenamefont {Flacke},
  \citenamefont {Gonzalez-Fraile}, \citenamefont {Lee}, \citenamefont {Panico}
  \emph {et~al.}}]{Delaunay:2013pwa}%
  \BibitemOpen
  \bibfield  {author} {\bibinfo {author} {\bibfnamefont {C.}~\bibnamefont
  {Delaunay}}, \bibinfo {author} {\bibfnamefont {T.}~\bibnamefont {Flacke}},
  \bibinfo {author} {\bibfnamefont {J.}~\bibnamefont {Gonzalez-Fraile}},
  \bibinfo {author} {\bibfnamefont {S.~J.}\ \bibnamefont {Lee}}, \bibinfo
  {author} {\bibfnamefont {G.}~\bibnamefont {Panico}},  \emph {et~al.},\ }\href
  {\doibase 10.1007/JHEP02(2014)055} {\bibfield  {journal} {\bibinfo  {journal}
  {JHEP}\ }\textbf {\bibinfo {volume} {1402}},\ \bibinfo {pages} {055}
  (\bibinfo {year} {2014}{\natexlab{a}})},\ \Eprint
  {http://arxiv.org/abs/1311.2072} {arXiv:1311.2072 [hep-ph]} \BibitemShut
  {NoStop}%
\bibitem [{\citenamefont {Blanke}\ \emph {et~al.}(2013)\citenamefont {Blanke},
  \citenamefont {Giudice}, \citenamefont {Paradisi}, \citenamefont {Perez},\
  and\ \citenamefont {Zupan}}]{Blanke:2013uia}%
  \BibitemOpen
  \bibfield  {author} {\bibinfo {author} {\bibfnamefont {M.}~\bibnamefont
  {Blanke}}, \bibinfo {author} {\bibfnamefont {G.~F.}\ \bibnamefont {Giudice}},
  \bibinfo {author} {\bibfnamefont {P.}~\bibnamefont {Paradisi}}, \bibinfo
  {author} {\bibfnamefont {G.}~\bibnamefont {Perez}}, \ and\ \bibinfo {author}
  {\bibfnamefont {J.}~\bibnamefont {Zupan}},\ }\href {\doibase
  10.1007/JHEP06(2013)022} {\bibfield  {journal} {\bibinfo  {journal} {JHEP}\
  }\textbf {\bibinfo {volume} {1306}},\ \bibinfo {pages} {022} (\bibinfo {year}
  {2013})},\ \Eprint {http://arxiv.org/abs/1302.7232} {arXiv:1302.7232
  [hep-ph]} \BibitemShut {NoStop}%
\bibitem [{\citenamefont {Mahbubani}\ \emph {et~al.}(2013)\citenamefont
  {Mahbubani}, \citenamefont {Papucci}, \citenamefont {Perez}, \citenamefont
  {Ruderman},\ and\ \citenamefont {Weiler}}]{Mahbubani:2012qq}%
  \BibitemOpen
  \bibfield  {author} {\bibinfo {author} {\bibfnamefont {R.}~\bibnamefont
  {Mahbubani}}, \bibinfo {author} {\bibfnamefont {M.}~\bibnamefont {Papucci}},
  \bibinfo {author} {\bibfnamefont {G.}~\bibnamefont {Perez}}, \bibinfo
  {author} {\bibfnamefont {J.~T.}\ \bibnamefont {Ruderman}}, \ and\ \bibinfo
  {author} {\bibfnamefont {A.}~\bibnamefont {Weiler}},\ }\href {\doibase
  10.1103/PhysRevLett.110.151804} {\bibfield  {journal} {\bibinfo  {journal}
  {Phys.Rev.Lett.}\ }\textbf {\bibinfo {volume} {110}},\ \bibinfo {pages}
  {151804} (\bibinfo {year} {2013})},\ \Eprint {http://arxiv.org/abs/1212.3328}
  {arXiv:1212.3328 [hep-ph]} \BibitemShut {NoStop}%
\bibitem [{\citenamefont {Kagan}\ \emph {et~al.}(2009)\citenamefont {Kagan},
  \citenamefont {Perez}, \citenamefont {Volansky},\ and\ \citenamefont
  {Zupan}}]{Kagan:2009bn}%
  \BibitemOpen
  \bibfield  {author} {\bibinfo {author} {\bibfnamefont {A.~L.}\ \bibnamefont
  {Kagan}}, \bibinfo {author} {\bibfnamefont {G.}~\bibnamefont {Perez}},
  \bibinfo {author} {\bibfnamefont {T.}~\bibnamefont {Volansky}}, \ and\
  \bibinfo {author} {\bibfnamefont {J.}~\bibnamefont {Zupan}},\ }\href
  {\doibase 10.1103/PhysRevD.80.076002} {\bibfield  {journal} {\bibinfo
  {journal} {Phys.Rev.}\ }\textbf {\bibinfo {volume} {D80}},\ \bibinfo {pages}
  {076002} (\bibinfo {year} {2009})},\ \Eprint {http://arxiv.org/abs/0903.1794}
  {arXiv:0903.1794 [hep-ph]} \BibitemShut {NoStop}%
\bibitem [{\citenamefont {Dery}\ \emph {et~al.}(2013)\citenamefont {Dery},
  \citenamefont {Efrati}, \citenamefont {Hiller}, \citenamefont {Hochberg},\
  and\ \citenamefont {Nir}}]{Dery:2013aba}%
  \BibitemOpen
  \bibfield  {author} {\bibinfo {author} {\bibfnamefont {A.}~\bibnamefont
  {Dery}}, \bibinfo {author} {\bibfnamefont {A.}~\bibnamefont {Efrati}},
  \bibinfo {author} {\bibfnamefont {G.}~\bibnamefont {Hiller}}, \bibinfo
  {author} {\bibfnamefont {Y.}~\bibnamefont {Hochberg}}, \ and\ \bibinfo
  {author} {\bibfnamefont {Y.}~\bibnamefont {Nir}},\ }\href {\doibase
  10.1007/JHEP08(2013)006} {\bibfield  {journal} {\bibinfo  {journal} {JHEP}\
  }\textbf {\bibinfo {volume} {1308}},\ \bibinfo {pages} {006} (\bibinfo {year}
  {2013})},\ \Eprint {http://arxiv.org/abs/1304.6727} {arXiv:1304.6727}
  \BibitemShut {NoStop}%
\bibitem [{\citenamefont {Giudice}\ and\ \citenamefont
  {Lebedev}(2008)}]{Giudice:2008uua}%
  \BibitemOpen
  \bibfield  {author} {\bibinfo {author} {\bibfnamefont {G.~F.}\ \bibnamefont
  {Giudice}}\ and\ \bibinfo {author} {\bibfnamefont {O.}~\bibnamefont
  {Lebedev}},\ }\href {\doibase 10.1016/j.physletb.2008.05.062} {\bibfield
  {journal} {\bibinfo  {journal} {Phys.Lett.}\ }\textbf {\bibinfo {volume}
  {B665}},\ \bibinfo {pages} {79} (\bibinfo {year} {2008})},\ \Eprint
  {http://arxiv.org/abs/0804.1753} {arXiv:0804.1753 [hep-ph]} \BibitemShut
  {NoStop}%
\bibitem [{\citenamefont {Da~Rold}\ \emph {et~al.}(2013)\citenamefont
  {Da~Rold}, \citenamefont {Delaunay}, \citenamefont {Grojean},\ and\
  \citenamefont {Perez}}]{DaRold:2012sz}%
  \BibitemOpen
  \bibfield  {author} {\bibinfo {author} {\bibfnamefont {L.}~\bibnamefont
  {Da~Rold}}, \bibinfo {author} {\bibfnamefont {C.}~\bibnamefont {Delaunay}},
  \bibinfo {author} {\bibfnamefont {C.}~\bibnamefont {Grojean}}, \ and\
  \bibinfo {author} {\bibfnamefont {G.}~\bibnamefont {Perez}},\ }\href
  {\doibase 10.1007/JHEP02(2013)149} {\bibfield  {journal} {\bibinfo  {journal}
  {JHEP}\ }\textbf {\bibinfo {volume} {1302}},\ \bibinfo {pages} {149}
  (\bibinfo {year} {2013})},\ \Eprint {http://arxiv.org/abs/1208.1499}
  {arXiv:1208.1499 [hep-ph]} \BibitemShut {NoStop}%
\bibitem [{\citenamefont {Chen}\ \emph {et~al.}(2013)\citenamefont {Chen},
  \citenamefont {Hou}, \citenamefont {Kao},\ and\ \citenamefont
  {Kohda}}]{Chen:2013qta}%
  \BibitemOpen
  \bibfield  {author} {\bibinfo {author} {\bibfnamefont {K.-F.}\ \bibnamefont
  {Chen}}, \bibinfo {author} {\bibfnamefont {W.-S.}\ \bibnamefont {Hou}},
  \bibinfo {author} {\bibfnamefont {C.}~\bibnamefont {Kao}}, \ and\ \bibinfo
  {author} {\bibfnamefont {M.}~\bibnamefont {Kohda}},\ }\href {\doibase
  10.1016/j.physletb.2013.07.060} {\bibfield  {journal} {\bibinfo  {journal}
  {Phys.Lett.}\ }\textbf {\bibinfo {volume} {B725}},\ \bibinfo {pages} {378}
  (\bibinfo {year} {2013})},\ \Eprint {http://arxiv.org/abs/1304.8037}
  {arXiv:1304.8037 [hep-ph]} \BibitemShut {NoStop}%
\bibitem [{\citenamefont {Dery}\ \emph {et~al.}(2014)\citenamefont {Dery},
  \citenamefont {Efrati}, \citenamefont {Nir}, \citenamefont {Soreq},\ and\
  \citenamefont {Susi\u{c}}}]{Dery:2014kxa}%
  \BibitemOpen
  \bibfield  {author} {\bibinfo {author} {\bibfnamefont {A.}~\bibnamefont
  {Dery}}, \bibinfo {author} {\bibfnamefont {A.}~\bibnamefont {Efrati}},
  \bibinfo {author} {\bibfnamefont {Y.}~\bibnamefont {Nir}}, \bibinfo {author}
  {\bibfnamefont {Y.}~\bibnamefont {Soreq}}, \ and\ \bibinfo {author}
  {\bibfnamefont {V.}~\bibnamefont {Susi\u{c}}},\ }\href {\doibase
  10.1103/PhysRevD.90.115022} {\bibfield  {journal} {\bibinfo  {journal}
  {Phys.Rev.}\ }\textbf {\bibinfo {volume} {D90}},\ \bibinfo {pages} {115022}
  (\bibinfo {year} {2014})},\ \Eprint {http://arxiv.org/abs/1408.1371}
  {arXiv:1408.1371 [hep-ph]} \BibitemShut {NoStop}%
\bibitem [{\citenamefont {Delaunay}\ \emph
  {et~al.}(2014{\natexlab{b}})\citenamefont {Delaunay}, \citenamefont
  {Golling}, \citenamefont {Perez},\ and\ \citenamefont
  {Soreq}}]{Delaunay:2013pja}%
  \BibitemOpen
  \bibfield  {author} {\bibinfo {author} {\bibfnamefont {C.}~\bibnamefont
  {Delaunay}}, \bibinfo {author} {\bibfnamefont {T.}~\bibnamefont {Golling}},
  \bibinfo {author} {\bibfnamefont {G.}~\bibnamefont {Perez}}, \ and\ \bibinfo
  {author} {\bibfnamefont {Y.}~\bibnamefont {Soreq}},\ }\href {\doibase
  10.1103/PhysRevD.89.033014} {\bibfield  {journal} {\bibinfo  {journal}
  {Phys.Rev.}\ }\textbf {\bibinfo {volume} {D89}},\ \bibinfo {pages} {033014}
  (\bibinfo {year} {2014}{\natexlab{b}})},\ \Eprint
  {http://arxiv.org/abs/1310.7029} {arXiv:1310.7029 [hep-ph]} \BibitemShut
  {NoStop}%
\bibitem [{\citenamefont {Bodwin}\ \emph {et~al.}(2013)\citenamefont {Bodwin},
  \citenamefont {Petriello}, \citenamefont {Stoynev},\ and\ \citenamefont
  {Velasco}}]{Bodwin:2013gca}%
  \BibitemOpen
  \bibfield  {author} {\bibinfo {author} {\bibfnamefont {G.~T.}\ \bibnamefont
  {Bodwin}}, \bibinfo {author} {\bibfnamefont {F.}~\bibnamefont {Petriello}},
  \bibinfo {author} {\bibfnamefont {S.}~\bibnamefont {Stoynev}}, \ and\
  \bibinfo {author} {\bibfnamefont {M.}~\bibnamefont {Velasco}},\ }\href
  {\doibase 10.1103/PhysRevD.88.053003} {\bibfield  {journal} {\bibinfo
  {journal} {Phys.Rev.}\ }\textbf {\bibinfo {volume} {D88}},\ \bibinfo {pages}
  {053003} (\bibinfo {year} {2013})},\ \Eprint {http://arxiv.org/abs/1306.5770}
  {arXiv:1306.5770 [hep-ph]} \BibitemShut {NoStop}%
\bibitem [{\citenamefont {Kagan}\ \emph {et~al.}(2014)\citenamefont {Kagan},
  \citenamefont {Perez}, \citenamefont {Petriello}, \citenamefont {Soreq},
  \citenamefont {Stoynev} \emph {et~al.}}]{Kagan:2014ila}%
  \BibitemOpen
  \bibfield  {author} {\bibinfo {author} {\bibfnamefont {A.~L.}\ \bibnamefont
  {Kagan}}, \bibinfo {author} {\bibfnamefont {G.}~\bibnamefont {Perez}},
  \bibinfo {author} {\bibfnamefont {F.}~\bibnamefont {Petriello}}, \bibinfo
  {author} {\bibfnamefont {Y.}~\bibnamefont {Soreq}}, \bibinfo {author}
  {\bibfnamefont {S.}~\bibnamefont {Stoynev}},  \emph {et~al.},\ }\href@noop {}
  {\  (\bibinfo {year} {2014})},\ \Eprint {http://arxiv.org/abs/1406.1722}
  {arXiv:1406.1722 [hep-ph]} \BibitemShut {NoStop}%
\bibitem [{\citenamefont {Mangano}\ and\ \citenamefont
  {Melia}(2014)}]{Mangano:2014xta}%
  \BibitemOpen
  \bibfield  {author} {\bibinfo {author} {\bibfnamefont {M.}~\bibnamefont
  {Mangano}}\ and\ \bibinfo {author} {\bibfnamefont {T.}~\bibnamefont
  {Melia}},\ }\href@noop {} {\  (\bibinfo {year} {2014})},\ \Eprint
  {http://arxiv.org/abs/1410.7475} {arXiv:1410.7475 [hep-ph]} \BibitemShut
  {NoStop}%
\bibitem [{\citenamefont {Huang}\ and\ \citenamefont
  {Petriello}(2014)}]{Huang:2014cxa}%
  \BibitemOpen
  \bibfield  {author} {\bibinfo {author} {\bibfnamefont {T.-C.}\ \bibnamefont
  {Huang}}\ and\ \bibinfo {author} {\bibfnamefont {F.}~\bibnamefont
  {Petriello}},\ }\href@noop {} {\  (\bibinfo {year} {2014})},\ \Eprint
  {http://arxiv.org/abs/1411.5924} {arXiv:1411.5924 [hep-ph]} \BibitemShut
  {NoStop}%
\bibitem [{\citenamefont {Grossman}\ \emph {et~al.}(2015)\citenamefont
  {Grossman}, \citenamefont {K{\"o}nigÂ},\ and\ \citenamefont
  {Neubert}}]{Grossmann:2015lea}%
  \BibitemOpen
  \bibfield  {author} {\bibinfo {author} {\bibfnamefont {Y.}~\bibnamefont
  {Grossman}}, \bibinfo {author} {\bibfnamefont {M.}~\bibnamefont
  {K{\"o}nigÂ}}, \ and\ \bibinfo {author} {\bibfnamefont {M.}~\bibnamefont
  {Neubert}},\ }\href@noop {} {\  (\bibinfo {year} {2015})},\ \Eprint
  {http://arxiv.org/abs/1501.06569} {arXiv:1501.06569 [hep-ph]} \BibitemShut
  {NoStop}%
\bibitem [{\citenamefont {Aad}\ \emph {et~al.}(2014{\natexlab{c}})\citenamefont
  {Aad} \emph {et~al.}}]{Aad:2014nra}%
  \BibitemOpen
  \bibfield  {author} {\bibinfo {author} {\bibfnamefont {G.}~\bibnamefont
  {Aad}} \emph {et~al.} (\bibinfo {collaboration} {ATLAS Collaboration}),\
  }\href {\doibase 10.1103/PhysRevD.90.052008} {\bibfield  {journal} {\bibinfo
  {journal} {Phys.Rev.}\ }\textbf {\bibinfo {volume} {D90}},\ \bibinfo {pages}
  {052008} (\bibinfo {year} {2014}{\natexlab{c}})},\ \Eprint
  {http://arxiv.org/abs/1407.0608} {arXiv:1407.0608 [hep-ex]} \BibitemShut
  {NoStop}%
\bibitem [{\citenamefont {Aad}\ \emph {et~al.}(2015{\natexlab{b}})\citenamefont
  {Aad} \emph {et~al.}}]{Aad:2015gna}%
  \BibitemOpen
  \bibfield  {author} {\bibinfo {author} {\bibfnamefont {G.}~\bibnamefont
  {Aad}} \emph {et~al.} (\bibinfo {collaboration} {ATLAS Collaboration}),\
  }\href@noop {} {\  (\bibinfo {year} {2015}{\natexlab{b}})},\ \Eprint
  {http://arxiv.org/abs/1501.01325} {arXiv:1501.01325 [hep-ex]} \BibitemShut
  {NoStop}%
\bibitem [{ATL(2015)}]{ATL-PHYS-PUB-2015-001}%
  \BibitemOpen
  \href@noop {} {\emph {\bibinfo {title} {{Performance and Calibration of the
  JetFitterCharm Algorithm for c-Jet Identification}}}},\ \bibinfo {type}
  {Tech. Rep.}\ \bibinfo {number} {ATL-PHYS-PUB-2015-001}\ (\bibinfo
  {institution} {CERN},\ \bibinfo {address} {Geneva},\ \bibinfo {year}
  {2015})\BibitemShut {NoStop}%
\bibitem [{\citenamefont {Aad}\ \emph {et~al.}(2015{\natexlab{c}})\citenamefont
  {Aad} \emph {et~al.}}]{Aad:2015sda}%
  \BibitemOpen
  \bibfield  {author} {\bibinfo {author} {\bibfnamefont {G.}~\bibnamefont
  {Aad}} \emph {et~al.} (\bibinfo {collaboration} {ATLAS Collaboration}),\
  }\href@noop {} {\  (\bibinfo {year} {2015}{\natexlab{c}})},\ \Eprint
  {http://arxiv.org/abs/1501.03276} {arXiv:1501.03276 [hep-ex]} \BibitemShut
  {NoStop}%
\bibitem [{\citenamefont {Heinemeyer}\ \emph {et~al.}(2013)\citenamefont
  {Heinemeyer} \emph {et~al.}}]{Heinemeyer:2013tqa}%
  \BibitemOpen
  \bibfield  {author} {\bibinfo {author} {\bibfnamefont {S.}~\bibnamefont
  {Heinemeyer}} \emph {et~al.} (\bibinfo {collaboration} {LHC Higgs Cross
  Section Working Group}),\ }\href {\doibase 10.5170/CERN-2013-004} {\
  (\bibinfo {year} {2013}),\ 10.5170/CERN-2013-004},\ \Eprint
  {http://arxiv.org/abs/1307.1347} {arXiv:1307.1347 [hep-ph]} \BibitemShut
  {NoStop}%
\bibitem [{The ATLAS Collaboration, ATLAS-CONF-2014-046,
  ATLAS-COM-CONF-2013-087()}]{ATLASbtag}%
  \BibitemOpen
  \bibfield  {author} {The ATLAS Collaboration, ATLAS-CONF-2014-046,
  ATLAS-COM-CONF-2013-087,\ }\href@noop {} {\ }\BibitemShut {NoStop}%
\bibitem [{\citenamefont {Chatrchyan}\ \emph {et~al.}(2013)\citenamefont
  {Chatrchyan} \emph {et~al.}}]{Chatrchyan:2012jua}%
  \BibitemOpen
  \bibfield  {author} {\bibinfo {author} {\bibfnamefont {S.}~\bibnamefont
  {Chatrchyan}} \emph {et~al.} (\bibinfo {collaboration} {CMS Collaboration}),\
  }\href {\doibase 10.1088/1748-0221/8/04/P04013} {\bibfield  {journal}
  {\bibinfo  {journal} {JINST}\ }\textbf {\bibinfo {volume} {8}},\ \bibinfo
  {pages} {P04013} (\bibinfo {year} {2013})},\ \Eprint
  {http://arxiv.org/abs/1211.4462} {arXiv:1211.4462 [hep-ex]} \BibitemShut
  {NoStop}%
\bibitem [{\citenamefont {Cowan}\ \emph {et~al.}(2011)\citenamefont {Cowan},
  \citenamefont {Cranmer}, \citenamefont {Gross},\ and\ \citenamefont
  {Vitells}}]{Cowan:2010js}%
  \BibitemOpen
  \bibfield  {author} {\bibinfo {author} {\bibfnamefont {G.}~\bibnamefont
  {Cowan}}, \bibinfo {author} {\bibfnamefont {K.}~\bibnamefont {Cranmer}},
  \bibinfo {author} {\bibfnamefont {E.}~\bibnamefont {Gross}}, \ and\ \bibinfo
  {author} {\bibfnamefont {O.}~\bibnamefont {Vitells}},\ }\href {\doibase
  10.1140/epjc/s10052-011-1554-0, 10.1140/epjc/s10052-013-2501-z} {\bibfield
  {journal} {\bibinfo  {journal} {Eur.Phys.J.}\ }\textbf {\bibinfo {volume}
  {C71}},\ \bibinfo {pages} {1554} (\bibinfo {year} {2011})},\ \Eprint
  {http://arxiv.org/abs/1007.1727} {arXiv:1007.1727 [physics.data-an]}
  \BibitemShut {NoStop}%
\bibitem [{\citenamefont {Alwall}\ \emph {et~al.}(2011)\citenamefont {Alwall},
  \citenamefont {Herquet}, \citenamefont {Maltoni}, \citenamefont {Mattelaer},\
  and\ \citenamefont {Stelzer}}]{Alwall:2011uj}%
  \BibitemOpen
  \bibfield  {author} {\bibinfo {author} {\bibfnamefont {J.}~\bibnamefont
  {Alwall}}, \bibinfo {author} {\bibfnamefont {M.}~\bibnamefont {Herquet}},
  \bibinfo {author} {\bibfnamefont {F.}~\bibnamefont {Maltoni}}, \bibinfo
  {author} {\bibfnamefont {O.}~\bibnamefont {Mattelaer}}, \ and\ \bibinfo
  {author} {\bibfnamefont {T.}~\bibnamefont {Stelzer}},\ }\href {\doibase
  10.1007/JHEP06(2011)128} {\bibfield  {journal} {\bibinfo  {journal} {JHEP}\
  }\textbf {\bibinfo {volume} {1106}},\ \bibinfo {pages} {128} (\bibinfo {year}
  {2011})},\ \Eprint {http://arxiv.org/abs/1106.0522} {arXiv:1106.0522
  [hep-ph]} \BibitemShut {NoStop}%
\bibitem [{\citenamefont {Perez}\ \emph {et~al.}(tion)\citenamefont {Perez},
  \citenamefont {Soreq}, \citenamefont {Stamou},\ and\ \citenamefont
  {Tobioka}}]{PSST}%
  \BibitemOpen
  \bibfield  {author} {\bibinfo {author} {\bibfnamefont {G.}~\bibnamefont
  {Perez}}, \bibinfo {author} {\bibfnamefont {Y.}~\bibnamefont {Soreq}},
  \bibinfo {author} {\bibfnamefont {E.}~\bibnamefont {Stamou}}, \ and\ \bibinfo
  {author} {\bibfnamefont {K.}~\bibnamefont {Tobioka}},\ }\href@noop {} {}
  (\bibinfo {year} {in preparation})\BibitemShut {NoStop}%
\bibitem [{\citenamefont {Aad}\ \emph {et~al.}(2014{\natexlab{d}})\citenamefont
  {Aad} \emph {et~al.}}]{Aad:2014aba}%
  \BibitemOpen
  \bibfield  {author} {\bibinfo {author} {\bibfnamefont {G.}~\bibnamefont
  {Aad}} \emph {et~al.} (\bibinfo {collaboration} {ATLAS Collaboration}),\
  }\href {\doibase 10.1103/PhysRevD.90.052004} {\bibfield  {journal} {\bibinfo
  {journal} {Phys.Rev.}\ }\textbf {\bibinfo {volume} {D90}},\ \bibinfo {pages}
  {052004} (\bibinfo {year} {2014}{\natexlab{d}})},\ \Eprint
  {http://arxiv.org/abs/1406.3827} {arXiv:1406.3827 [hep-ex]} \BibitemShut
  {NoStop}%
\bibitem [{\citenamefont {Khachatryan}\ \emph
  {et~al.}(2014{\natexlab{c}})\citenamefont {Khachatryan} \emph
  {et~al.}}]{Khachatryan:2014jba}%
  \BibitemOpen
  \bibfield  {author} {\bibinfo {author} {\bibfnamefont {V.}~\bibnamefont
  {Khachatryan}} \emph {et~al.} (\bibinfo {collaboration} {CMS
  Collaboration}),\ }\href@noop {} {\  (\bibinfo {year}
  {2014}{\natexlab{c}})},\ \Eprint {http://arxiv.org/abs/1412.8662}
  {arXiv:1412.8662 [hep-ex]} \BibitemShut {NoStop}%
\bibitem [{\citenamefont {Bodwin}\ \emph {et~al.}(2014)\citenamefont {Bodwin},
  \citenamefont {Chung}, \citenamefont {Ee}, \citenamefont {Lee},\ and\
  \citenamefont {Petriello}}]{Bodwin:2014bpa}%
  \BibitemOpen
  \bibfield  {author} {\bibinfo {author} {\bibfnamefont {G.~T.}\ \bibnamefont
  {Bodwin}}, \bibinfo {author} {\bibfnamefont {H.~S.}\ \bibnamefont {Chung}},
  \bibinfo {author} {\bibfnamefont {J.-H.}\ \bibnamefont {Ee}}, \bibinfo
  {author} {\bibfnamefont {J.}~\bibnamefont {Lee}}, \ and\ \bibinfo {author}
  {\bibfnamefont {F.}~\bibnamefont {Petriello}},\ }\href {\doibase
  10.1103/PhysRevD.90.113010} {\bibfield  {journal} {\bibinfo  {journal}
  {Phys.Rev.}\ }\textbf {\bibinfo {volume} {D90}},\ \bibinfo {pages} {113010}
  (\bibinfo {year} {2014})},\ \Eprint {http://arxiv.org/abs/1407.6695}
  {arXiv:1407.6695 [hep-ph]} \BibitemShut {NoStop}%
\bibitem [{\citenamefont {Aad}\ \emph {et~al.}(2015{\natexlab{d}})\citenamefont
  {Aad} \emph {et~al.}}]{Aad:2014eva}%
  \BibitemOpen
  \bibfield  {author} {\bibinfo {author} {\bibfnamefont {G.}~\bibnamefont
  {Aad}} \emph {et~al.} (\bibinfo {collaboration} {ATLAS Collaboration}),\
  }\href {\doibase 10.1103/PhysRevD.91.012006} {\bibfield  {journal} {\bibinfo
  {journal} {Phys.Rev.}\ }\textbf {\bibinfo {volume} {D91}},\ \bibinfo {pages}
  {012006} (\bibinfo {year} {2015}{\natexlab{d}})},\ \Eprint
  {http://arxiv.org/abs/1408.5191} {arXiv:1408.5191 [hep-ex]} \BibitemShut
  {NoStop}%
\bibitem [{\citenamefont {Carmi}\ \emph {et~al.}(2012)\citenamefont {Carmi},
  \citenamefont {Falkowski}, \citenamefont {Kuflik}, \citenamefont {Volansky},\
  and\ \citenamefont {Zupan}}]{Carmi:2012in}%
  \BibitemOpen
  \bibfield  {author} {\bibinfo {author} {\bibfnamefont {D.}~\bibnamefont
  {Carmi}}, \bibinfo {author} {\bibfnamefont {A.}~\bibnamefont {Falkowski}},
  \bibinfo {author} {\bibfnamefont {E.}~\bibnamefont {Kuflik}}, \bibinfo
  {author} {\bibfnamefont {T.}~\bibnamefont {Volansky}}, \ and\ \bibinfo
  {author} {\bibfnamefont {J.}~\bibnamefont {Zupan}},\ }\href {\doibase
  10.1007/JHEP10(2012)196} {\bibfield  {journal} {\bibinfo  {journal} {JHEP}\
  }\textbf {\bibinfo {volume} {1210}},\ \bibinfo {pages} {196} (\bibinfo {year}
  {2012})},\ \Eprint {http://arxiv.org/abs/1207.1718} {arXiv:1207.1718
  [hep-ph]} \BibitemShut {NoStop}%
\bibitem [{\citenamefont {Giardino}\ \emph {et~al.}(2012)\citenamefont
  {Giardino}, \citenamefont {Kannike}, \citenamefont {Raidal},\ and\
  \citenamefont {Strumia}}]{Giardino:2012dp}%
  \BibitemOpen
  \bibfield  {author} {\bibinfo {author} {\bibfnamefont {P.~P.}\ \bibnamefont
  {Giardino}}, \bibinfo {author} {\bibfnamefont {K.}~\bibnamefont {Kannike}},
  \bibinfo {author} {\bibfnamefont {M.}~\bibnamefont {Raidal}}, \ and\ \bibinfo
  {author} {\bibfnamefont {A.}~\bibnamefont {Strumia}},\ }\href {\doibase
  10.1016/j.physletb.2012.10.042} {\bibfield  {journal} {\bibinfo  {journal}
  {Phys.Lett.}\ }\textbf {\bibinfo {volume} {B718}},\ \bibinfo {pages} {469}
  (\bibinfo {year} {2012})},\ \Eprint {http://arxiv.org/abs/1207.1347}
  {arXiv:1207.1347 [hep-ph]} \BibitemShut {NoStop}%
\bibitem [{\citenamefont {Espinosa}\ \emph
  {et~al.}(2012{\natexlab{a}})\citenamefont {Espinosa}, \citenamefont
  {Grojean}, \citenamefont {Muhlleitner},\ and\ \citenamefont
  {Trott}}]{Espinosa:2012im}%
  \BibitemOpen
  \bibfield  {author} {\bibinfo {author} {\bibfnamefont {J.}~\bibnamefont
  {Espinosa}}, \bibinfo {author} {\bibfnamefont {C.}~\bibnamefont {Grojean}},
  \bibinfo {author} {\bibfnamefont {M.}~\bibnamefont {Muhlleitner}}, \ and\
  \bibinfo {author} {\bibfnamefont {M.}~\bibnamefont {Trott}},\ }\href
  {\doibase 10.1007/JHEP12(2012)045} {\bibfield  {journal} {\bibinfo  {journal}
  {JHEP}\ }\textbf {\bibinfo {volume} {1212}},\ \bibinfo {pages} {045}
  (\bibinfo {year} {2012}{\natexlab{a}})},\ \Eprint
  {http://arxiv.org/abs/1207.1717} {arXiv:1207.1717 [hep-ph]} \BibitemShut
  {NoStop}%
\bibitem [{\citenamefont {Ellis}\ and\ \citenamefont
  {You}(2012)}]{Ellis:2012hz}%
  \BibitemOpen
  \bibfield  {author} {\bibinfo {author} {\bibfnamefont {J.}~\bibnamefont
  {Ellis}}\ and\ \bibinfo {author} {\bibfnamefont {T.}~\bibnamefont {You}},\
  }\href {\doibase 10.1007/JHEP09(2012)123} {\bibfield  {journal} {\bibinfo
  {journal} {JHEP}\ }\textbf {\bibinfo {volume} {1209}},\ \bibinfo {pages}
  {123} (\bibinfo {year} {2012})},\ \Eprint {http://arxiv.org/abs/1207.1693}
  {arXiv:1207.1693 [hep-ph]} \BibitemShut {NoStop}%
\bibitem [{\citenamefont {Falkowski}\ \emph {et~al.}(2013)\citenamefont
  {Falkowski}, \citenamefont {Riva},\ and\ \citenamefont
  {Urbano}}]{Falkowski:2013dza}%
  \BibitemOpen
  \bibfield  {author} {\bibinfo {author} {\bibfnamefont {A.}~\bibnamefont
  {Falkowski}}, \bibinfo {author} {\bibfnamefont {F.}~\bibnamefont {Riva}}, \
  and\ \bibinfo {author} {\bibfnamefont {A.}~\bibnamefont {Urbano}},\ }\href
  {\doibase 10.1007/JHEP11(2013)111} {\bibfield  {journal} {\bibinfo  {journal}
  {JHEP}\ }\textbf {\bibinfo {volume} {1311}},\ \bibinfo {pages} {111}
  (\bibinfo {year} {2013})},\ \Eprint {http://arxiv.org/abs/1303.1812}
  {arXiv:1303.1812 [hep-ph]} \BibitemShut {NoStop}%
\bibitem [{\citenamefont {Espinosa}\ \emph
  {et~al.}(2012{\natexlab{b}})\citenamefont {Espinosa}, \citenamefont
  {Muhlleitner}, \citenamefont {Grojean},\ and\ \citenamefont
  {Trott}}]{Espinosa:2012vu}%
  \BibitemOpen
  \bibfield  {author} {\bibinfo {author} {\bibfnamefont {J.~R.}\ \bibnamefont
  {Espinosa}}, \bibinfo {author} {\bibfnamefont {M.}~\bibnamefont
  {Muhlleitner}}, \bibinfo {author} {\bibfnamefont {C.}~\bibnamefont
  {Grojean}}, \ and\ \bibinfo {author} {\bibfnamefont {M.}~\bibnamefont
  {Trott}},\ }\href {\doibase 10.1007/JHEP09(2012)126} {\bibfield  {journal}
  {\bibinfo  {journal} {JHEP}\ }\textbf {\bibinfo {volume} {1209}},\ \bibinfo
  {pages} {126} (\bibinfo {year} {2012}{\natexlab{b}})},\ \Eprint
  {http://arxiv.org/abs/1205.6790} {arXiv:1205.6790 [hep-ph]} \BibitemShut
  {NoStop}%
\bibitem [{\citenamefont {Dobrescu}\ and\ \citenamefont
  {Lykken}(2013)}]{Dobrescu:2012td}%
  \BibitemOpen
  \bibfield  {author} {\bibinfo {author} {\bibfnamefont {B.~A.}\ \bibnamefont
  {Dobrescu}}\ and\ \bibinfo {author} {\bibfnamefont {J.~D.}\ \bibnamefont
  {Lykken}},\ }\href {\doibase 10.1007/JHEP02(2013)073} {\bibfield  {journal}
  {\bibinfo  {journal} {JHEP}\ }\textbf {\bibinfo {volume} {1302}},\ \bibinfo
  {pages} {073} (\bibinfo {year} {2013})},\ \Eprint
  {http://arxiv.org/abs/1210.3342} {arXiv:1210.3342 [hep-ph]} \BibitemShut
  {NoStop}%
\bibitem [{\citenamefont {Bechtle}\ \emph
  {et~al.}(2014{\natexlab{a}})\citenamefont {Bechtle}, \citenamefont
  {Heinemeyer}, \citenamefont {St\r{a}l}, \citenamefont {Stefaniak},\ and\
  \citenamefont {Weiglein}}]{Bechtle:2014ewa}%
  \BibitemOpen
  \bibfield  {author} {\bibinfo {author} {\bibfnamefont {P.}~\bibnamefont
  {Bechtle}}, \bibinfo {author} {\bibfnamefont {S.}~\bibnamefont {Heinemeyer}},
  \bibinfo {author} {\bibfnamefont {O.}~\bibnamefont {St\r{a}l}}, \bibinfo
  {author} {\bibfnamefont {T.}~\bibnamefont {Stefaniak}}, \ and\ \bibinfo
  {author} {\bibfnamefont {G.}~\bibnamefont {Weiglein}},\ }\href {\doibase
  10.1007/JHEP11(2014)039} {\bibfield  {journal} {\bibinfo  {journal} {JHEP}\
  }\textbf {\bibinfo {volume} {1411}},\ \bibinfo {pages} {039} (\bibinfo {year}
  {2014}{\natexlab{a}})},\ \Eprint {http://arxiv.org/abs/1403.1582}
  {arXiv:1403.1582 [hep-ph]} \BibitemShut {NoStop}%
\bibitem [{\citenamefont {Aad}\ \emph {et~al.}(2014{\natexlab{e}})\citenamefont
  {Aad} \emph {et~al.}}]{Aad:2014eha}%
  \BibitemOpen
  \bibfield  {author} {\bibinfo {author} {\bibfnamefont {G.}~\bibnamefont
  {Aad}} \emph {et~al.} (\bibinfo {collaboration} {ATLAS Collaboration}),\
  }\href {\doibase 10.1103/PhysRevD.90.112015} {\bibfield  {journal} {\bibinfo
  {journal} {Phys.Rev.}\ }\textbf {\bibinfo {volume} {D90}},\ \bibinfo {pages}
  {112015} (\bibinfo {year} {2014}{\natexlab{e}})},\ \Eprint
  {http://arxiv.org/abs/1408.7084} {arXiv:1408.7084 [hep-ex]} \BibitemShut
  {NoStop}%
\bibitem [{The ATLAS Collaboration, ATLAS-CONF-2013-075,
  ATLAS-COM-CONF-2013-069()}]{TheATLAScollaboration:2013hia}%
  \BibitemOpen
  \bibfield  {author} {The ATLAS Collaboration, ATLAS-CONF-2013-075,
  ATLAS-COM-CONF-2013-069,\ }\href@noop {} {\  (\bibinfo {year}
  {2013})}\BibitemShut {NoStop}%
\bibitem [{The ATLAS Collaboration, ATLAS-CONF-2014-060,
  ATLAS-COM-CONF-2014-078()}]{ATLAS060}%
  \BibitemOpen
  \bibfield  {author} {The ATLAS Collaboration, ATLAS-CONF-2014-060,
  ATLAS-COM-CONF-2014-078,\ }\href@noop {} {\  (\bibinfo {year}
  {2014})}\BibitemShut {NoStop}%
\bibitem [{The ATLAS Collaboration, ATLAS-CONF-2014-061,
  ATLAS-COM-CONF-2014-080()}]{ATLAS061}%
  \BibitemOpen
  \bibfield  {author} {The ATLAS Collaboration, ATLAS-CONF-2014-061,
  ATLAS-COM-CONF-2014-080,\ }\href@noop {} {\  (\bibinfo {year}
  {2014})}\BibitemShut {NoStop}%
\bibitem [{\citenamefont {Chatrchyan}\ \emph
  {et~al.}(2014{\natexlab{c}})\citenamefont {Chatrchyan} \emph
  {et~al.}}]{Chatrchyan:2013iaa}%
  \BibitemOpen
  \bibfield  {author} {\bibinfo {author} {\bibfnamefont {S.}~\bibnamefont
  {Chatrchyan}} \emph {et~al.} (\bibinfo {collaboration} {CMS Collaboration}),\
  }\href {\doibase 10.1007/JHEP01(2014)096} {\bibfield  {journal} {\bibinfo
  {journal} {JHEP}\ }\textbf {\bibinfo {volume} {1401}},\ \bibinfo {pages}
  {096} (\bibinfo {year} {2014}{\natexlab{c}})},\ \Eprint
  {http://arxiv.org/abs/1312.1129} {arXiv:1312.1129 [hep-ex]} \BibitemShut
  {NoStop}%
\bibitem [{\citenamefont {Chatrchyan}\ \emph
  {et~al.}(2014{\natexlab{d}})\citenamefont {Chatrchyan} \emph
  {et~al.}}]{Chatrchyan:2013mxa}%
  \BibitemOpen
  \bibfield  {author} {\bibinfo {author} {\bibfnamefont {S.}~\bibnamefont
  {Chatrchyan}} \emph {et~al.} (\bibinfo {collaboration} {CMS Collaboration}),\
  }\href {\doibase 10.1103/PhysRevD.89.092007} {\bibfield  {journal} {\bibinfo
  {journal} {Phys.Rev.}\ }\textbf {\bibinfo {volume} {D89}},\ \bibinfo {pages}
  {092007} (\bibinfo {year} {2014}{\natexlab{d}})},\ \Eprint
  {http://arxiv.org/abs/1312.5353} {arXiv:1312.5353 [hep-ex]} \BibitemShut
  {NoStop}%
\bibitem [{\citenamefont {Khachatryan}\ \emph
  {et~al.}(2014{\natexlab{d}})\citenamefont {Khachatryan} \emph
  {et~al.}}]{Khachatryan:2014ira}%
  \BibitemOpen
  \bibfield  {author} {\bibinfo {author} {\bibfnamefont {V.}~\bibnamefont
  {Khachatryan}} \emph {et~al.} (\bibinfo {collaboration} {CMS
  Collaboration}),\ }\href {\doibase 10.1140/epjc/s10052-014-3076-z} {\bibfield
   {journal} {\bibinfo  {journal} {Eur.Phys.J.}\ }\textbf {\bibinfo {volume}
  {C74}},\ \bibinfo {pages} {3076} (\bibinfo {year} {2014}{\natexlab{d}})},\
  \Eprint {http://arxiv.org/abs/1407.0558} {arXiv:1407.0558 [hep-ex]}
  \BibitemShut {NoStop}%
\bibitem [{The CMS Collaboration, CMS-PAS-HIG-13-017()}]{CMS:2013xda}%
  \BibitemOpen
  \bibfield  {author} {The CMS Collaboration, CMS-PAS-HIG-13-017,\ }\href@noop
  {} {\  (\bibinfo {year} {2013})}\BibitemShut {NoStop}%
\bibitem [{\citenamefont {Aaltonen}\ \emph {et~al.}(2013)\citenamefont
  {Aaltonen} \emph {et~al.}}]{Aaltonen:2013ipa}%
  \BibitemOpen
  \bibfield  {author} {\bibinfo {author} {\bibfnamefont {T.}~\bibnamefont
  {Aaltonen}} \emph {et~al.} (\bibinfo {collaboration} {CDF Collaboration}),\
  }\href {\doibase 10.1103/PhysRevD.88.052013} {\bibfield  {journal} {\bibinfo
  {journal} {Phys.Rev.}\ }\textbf {\bibinfo {volume} {D88}},\ \bibinfo {pages}
  {052013} (\bibinfo {year} {2013})},\ \Eprint {http://arxiv.org/abs/1301.6668}
  {arXiv:1301.6668 [hep-ex]} \BibitemShut {NoStop}%
\bibitem [{\citenamefont {Abazov}\ \emph {et~al.}(2013)\citenamefont {Abazov}
  \emph {et~al.}}]{Abazov:2013gmz}%
  \BibitemOpen
  \bibfield  {author} {\bibinfo {author} {\bibfnamefont {V.~M.}\ \bibnamefont
  {Abazov}} \emph {et~al.} (\bibinfo {collaboration} {D0 Collaboration}),\
  }\href {\doibase 10.1103/PhysRevD.88.052011} {\bibfield  {journal} {\bibinfo
  {journal} {Phys.Rev.}\ }\textbf {\bibinfo {volume} {D88}},\ \bibinfo {pages}
  {052011} (\bibinfo {year} {2013})},\ \Eprint {http://arxiv.org/abs/1303.0823}
  {arXiv:1303.0823 [hep-ex]} \BibitemShut {NoStop}%
\bibitem [{\citenamefont {Bechtle}\ \emph
  {et~al.}(2014{\natexlab{b}})\citenamefont {Bechtle}, \citenamefont
  {Heinemeyer}, \citenamefont {St\r{a}l}, \citenamefont {Stefaniak},\ and\
  \citenamefont {Weiglein}}]{Bechtle:2013xfa}%
  \BibitemOpen
  \bibfield  {author} {\bibinfo {author} {\bibfnamefont {P.}~\bibnamefont
  {Bechtle}}, \bibinfo {author} {\bibfnamefont {S.}~\bibnamefont {Heinemeyer}},
  \bibinfo {author} {\bibfnamefont {O.}~\bibnamefont {St\r{a}l}}, \bibinfo
  {author} {\bibfnamefont {T.}~\bibnamefont {Stefaniak}}, \ and\ \bibinfo
  {author} {\bibfnamefont {G.}~\bibnamefont {Weiglein}},\ }\href {\doibase
  10.1140/epjc/s10052-013-2711-4} {\bibfield  {journal} {\bibinfo  {journal}
  {Eur.Phys.J.}\ }\textbf {\bibinfo {volume} {C74}},\ \bibinfo {pages} {2711}
  (\bibinfo {year} {2014}{\natexlab{b}})},\ \Eprint
  {http://arxiv.org/abs/1305.1933} {arXiv:1305.1933 [hep-ph]} \BibitemShut
  {NoStop}%
\bibitem [{\citenamefont {Caola}\ and\ \citenamefont
  {Melnikov}(2013)}]{Caola:2013yja}%
  \BibitemOpen
  \bibfield  {author} {\bibinfo {author} {\bibfnamefont {F.}~\bibnamefont
  {Caola}}\ and\ \bibinfo {author} {\bibfnamefont {K.}~\bibnamefont
  {Melnikov}},\ }\href {\doibase 10.1103/PhysRevD.88.054024} {\bibfield
  {journal} {\bibinfo  {journal} {Phys.Rev.}\ }\textbf {\bibinfo {volume}
  {D88}},\ \bibinfo {pages} {054024} (\bibinfo {year} {2013})},\ \Eprint
  {http://arxiv.org/abs/1307.4935} {arXiv:1307.4935 [hep-ph]} \BibitemShut
  {NoStop}%
\bibitem [{\citenamefont {Khachatryan}\ \emph
  {et~al.}(2014{\natexlab{e}})\citenamefont {Khachatryan} \emph
  {et~al.}}]{Khachatryan:2014iha}%
  \BibitemOpen
  \bibfield  {author} {\bibinfo {author} {\bibfnamefont {V.}~\bibnamefont
  {Khachatryan}} \emph {et~al.} (\bibinfo {collaboration} {CMS
  Collaboration}),\ }\href {\doibase 10.1016/j.physletb.2014.06.077} {\bibfield
   {journal} {\bibinfo  {journal} {Phys.Lett.}\ }\textbf {\bibinfo {volume}
  {B736}},\ \bibinfo {pages} {64} (\bibinfo {year} {2014}{\natexlab{e}})},\
  \Eprint {http://arxiv.org/abs/1405.3455} {arXiv:1405.3455 [hep-ex]}
  \BibitemShut {NoStop}%
\bibitem [{The ATLAS collaboration, ATLAS-CONF-2014-042,
  ATLAS-COM-CONF-2014-052()}]{ATLASoffshell}%
  \BibitemOpen
  \bibfield  {author} {The ATLAS collaboration, ATLAS-CONF-2014-042,
  ATLAS-COM-CONF-2014-052,\ }\href@noop {} {\  (\bibinfo {year}
  {2014})}\BibitemShut {NoStop}%
\bibitem [{\citenamefont {Englert}\ and\ \citenamefont
  {Spannowsky}(2014)}]{Englert:2014aca}%
  \BibitemOpen
  \bibfield  {author} {\bibinfo {author} {\bibfnamefont {C.}~\bibnamefont
  {Englert}}\ and\ \bibinfo {author} {\bibfnamefont {M.}~\bibnamefont
  {Spannowsky}},\ }\href {\doibase 10.1103/PhysRevD.90.053003} {\bibfield
  {journal} {\bibinfo  {journal} {Phys.Rev.}\ }\textbf {\bibinfo {volume}
  {D90}},\ \bibinfo {pages} {053003} (\bibinfo {year} {2014})},\ \Eprint
  {http://arxiv.org/abs/1405.0285} {arXiv:1405.0285 [hep-ph]} \BibitemShut
  {NoStop}%
\bibitem [{\citenamefont {Isidori}\ \emph {et~al.}(2014)\citenamefont
  {Isidori}, \citenamefont {Manohar},\ and\ \citenamefont
  {Trott}}]{Isidori:2013cla}%
  \BibitemOpen
  \bibfield  {author} {\bibinfo {author} {\bibfnamefont {G.}~\bibnamefont
  {Isidori}}, \bibinfo {author} {\bibfnamefont {A.~V.}\ \bibnamefont
  {Manohar}}, \ and\ \bibinfo {author} {\bibfnamefont {M.}~\bibnamefont
  {Trott}},\ }\href {\doibase 10.1016/j.physletb.2013.11.054} {\bibfield
  {journal} {\bibinfo  {journal} {Phys.Lett.}\ }\textbf {\bibinfo {volume}
  {B728}},\ \bibinfo {pages} {131} (\bibinfo {year} {2014})},\ \Eprint
  {http://arxiv.org/abs/1305.0663} {arXiv:1305.0663 [hep-ph]} \BibitemShut
  {NoStop}%
\bibitem [{\citenamefont {Harnik}\ \emph {et~al.}(2013)\citenamefont {Harnik},
  \citenamefont {Kopp},\ and\ \citenamefont {Zupan}}]{Harnik:2012pb}%
  \BibitemOpen
  \bibfield  {author} {\bibinfo {author} {\bibfnamefont {R.}~\bibnamefont
  {Harnik}}, \bibinfo {author} {\bibfnamefont {J.}~\bibnamefont {Kopp}}, \ and\
  \bibinfo {author} {\bibfnamefont {J.}~\bibnamefont {Zupan}},\ }\href
  {\doibase 10.1007/JHEP03(2013)026} {\bibfield  {journal} {\bibinfo  {journal}
  {JHEP}\ }\textbf {\bibinfo {volume} {1303}},\ \bibinfo {pages} {026}
  (\bibinfo {year} {2013})},\ \Eprint {http://arxiv.org/abs/1209.1397}
  {arXiv:1209.1397 [hep-ph]} \BibitemShut {NoStop}%
\bibitem [{\citenamefont {Goertz}(2014)}]{Goertz:2014qia}%
  \BibitemOpen
  \bibfield  {author} {\bibinfo {author} {\bibfnamefont {F.}~\bibnamefont
  {Goertz}},\ }\href {\doibase 10.1103/PhysRevLett.113.261803} {\bibfield
  {journal} {\bibinfo  {journal} {Phys.Rev.Lett.}\ }\textbf {\bibinfo {volume}
  {113}},\ \bibinfo {pages} {261803} (\bibinfo {year} {2014})},\ \Eprint
  {http://arxiv.org/abs/1406.0102} {arXiv:1406.0102 [hep-ph]} \BibitemShut
  {NoStop}%
\bibitem [{{The ATLAS Collaboration,
  ATL-PHYS-PUB-2014-011}()}]{ATL-PHYS-PUB-2014-011}%
  \BibitemOpen
  {The ATLAS Collaboration, ATL-PHYS-PUB-2014-011},\ \href@noop {} {\enquote
  {\bibinfo {title} {{Search for the $b\bar{b}$ decay of the Standard Model
  Higgs boson in associated $(W/Z)H$ production with the ATLAS detector}},}\ }
  (\bibinfo {year} {2014})\BibitemShut {NoStop}%
\bibitem [{\citenamefont {Capeans}\ \emph {et~al.}(2010)\citenamefont
  {Capeans}, \citenamefont {Darbo}, \citenamefont {Einsweiller}, \citenamefont
  {Elsing}, \citenamefont {Flick}, \citenamefont {Garcia-Sciveres},
  \citenamefont {Gemme}, \citenamefont {Pernegger}, \citenamefont {Rohne},\
  and\ \citenamefont {Vuillermet}}]{Capeans:1291633}%
  \BibitemOpen
  \bibfield  {author} {\bibinfo {author} {\bibfnamefont {M.}~\bibnamefont
  {Capeans}}, \bibinfo {author} {\bibfnamefont {G.}~\bibnamefont {Darbo}},
  \bibinfo {author} {\bibfnamefont {K.}~\bibnamefont {Einsweiller}}, \bibinfo
  {author} {\bibfnamefont {M.}~\bibnamefont {Elsing}}, \bibinfo {author}
  {\bibfnamefont {T.}~\bibnamefont {Flick}}, \bibinfo {author} {\bibfnamefont
  {M.}~\bibnamefont {Garcia-Sciveres}}, \bibinfo {author} {\bibfnamefont
  {C.}~\bibnamefont {Gemme}}, \bibinfo {author} {\bibfnamefont
  {H.}~\bibnamefont {Pernegger}}, \bibinfo {author} {\bibfnamefont
  {O.}~\bibnamefont {Rohne}}, \ and\ \bibinfo {author} {\bibfnamefont
  {R.}~\bibnamefont {Vuillermet}},\ }\href@noop {} {\emph {\bibinfo {title}
  {{ATLAS Insertable B-Layer Technical Design Report}}}},\ \bibinfo {type}
  {Tech. Rep.}\ \bibinfo {number} {CERN-LHCC-2010-013. ATLAS-TDR-19}\ (\bibinfo
   {institution} {CERN},\ \bibinfo {address} {Geneva},\ \bibinfo {year}
  {2010})\BibitemShut {NoStop}%
\end{thebibliography}%

\end{document}